\documentclass{aa}  

\usepackage{graphicx}
\usepackage{txfonts}
\usepackage{lipsum}
\usepackage{subcaption}
\usepackage{caption}  
\usepackage{lscape}  

\usepackage{placeins}  

\begin{document}

   \title{Interactions between the magnetised jets in\\ the twin radio galaxy J104455+354055}

   \author{M. We\.zgowiec\inst{1}
        \and R. Joshi\inst{2}
        \and Gopal-Krishna\inst{3}
        \and M. Jamrozy\inst{1}
        \and S. Mondal\inst{4}
        \and L.~C. Ho\inst{5,6}
        \and G. Giri\inst{7}
        }

   \institute{Obserwatorium Astronomiczne Uniwersytetu Jagiello\'nskiego, ul. Orla 171, 30-244 Krak\'ow, Poland\\
             \email{markmet@oa.uj.edu.pl}
            \and Indian Institute of Astrophysics (IIA), Koramangala, Bangalore 560034, India 
            \and UM-DAE Centre for Excellence in Basic Sciences, Mumbai, 400098, India
            \and Department of Physics and Electronics, Christ University, Hosur Main Road, Bangalore 560029, India
            \and Kavli Institute for Astronomy and Astrophysics, Peking University, Beijing 100871, China
            \and Department of Astronomy, School of Physics, Peking University, Beijing 100871, China
            \and Istituto Nazionale di Astrofisica (INAF) – Istituto di Radioastronomia (IRA), via Gobetti 101, 40129 Bologna, Italy}
   \date{}

 \titlerunning{Interactions between the magnetised jets...}
  \authorrunning{We\.zgowiec et al.}
 \abstract
   {
Pairs of galaxies in which both objects host kiloparsec-scale radio jets are extremely rare. Such systems 
provide unique opportunity to study the formation of relativistic jets and their resistance to
side-way collisions with each other prior to their eventual merger.
   }
   {
We analyse the radio data for the twin radio galaxy J104455+354055 associated with a pair of distant elliptical galaxies (z=0.163).
The radio jets of both sources avoid disruption/merging for the initial tens of kiloparsecs. We investigate the role of the magnetic 
fields in the propagation of these radio jets.
   }
   {
We used VLA radio data at high frequencies in both compact and extended array configurations 
to obtain sensitive high-resolution polarimetric data and study the properties of the magnetic fields in all jets. 
   }
   {
Polarisation data suggest shearing of the magnetic fields at points where the jets bend. The orientation of the magnetic field 
parallel to jet axis helps confine the jet. Such field configuration in two adjacent jets seems to prevent them from merging.
The strength of the magnetic field slightly increases in jet bends and the associated shearing enhances the ordered component 
of the magnetic field. The more diffuse radio plumes at the end of each jet have comparable strengths and energies of the magnetic field.
   }
   {
The twin radio galaxy J104455+354055 provides a unique possibility to study the interplay of radio jets of two radio galaxies. 
The analysis of the properties of the magnetic fields observed along the radio jets suggests that the wiggles and bends of 
these jets, caused by the likely interactions of the host galaxies, could be also governed by the magnetic fields that may prevent 
them from merging. The derived strengths and energies of the magnetic field in different parts of the jets should allow us to 
improve the magnetohydrodynamic (MHD) simulations of radio galaxies.
   }


   \maketitle
\nolinenumbers
\section{Introduction}
\label{intro}

Galaxy mergers are fundamental to the evolution of cosmic structures, leading to hierarchical galaxy formation
\citep[e.g.][]{conselice14, duncan19}. Observational evidence for this scenario comes from the abundance of pairs and groups of
galaxies  \citep[e.g.][]{menon85}. Pairs of similarly bright massive early-type galaxies have been referred to as ``dumb-bell'' galaxies \citep{harris60}.
A substantial number of dumb-bells is known, whereby one of the member galaxies hosts kiloparsec-scale (kpc-scale) radio jets. The kinematic effects of their orbital motion on the jets have been a topic of investigation since the pioneering work of \citet{wirth82}. Because nearly all large galaxies are believed to harbour a supermassive black hole (SMBH; \citealt{kormendy13}), we would expect to
find cases where both members of the galaxy pair eject powerful jets. Such pairs of twin radio jets, also called
twin radio galaxies (TRGs), can provide observational clues about the formation of jets and their resistance to side-way collisions with each other prior to their eventual merger. 
However, very few TRGs have been found so far. The prototype of this class, the radio source 3C\,75, extending over 0.5\,Mpc and located
in the galaxy cluster Abell\,400 (z=0.023), was discovered four decades ago \citep{owen85}. 
This TRG has also been a favourite target of numerical simulation studies \citep[e.g.][]{molnar17, musoke20}. 
The second TRG, discovered three decades ago, PKS\,2149$-$158 with a size of $\sim$400 kpc, is located near the outskirts 
of the cluster Abell\,2382 (z=0.062; \citealt{parma91, guidetti08}).

The third TRG known so far was reported very recently, underscoring the extreme rarity of TRGs \citep{gopal-krishna22}.
This source, J104455+354055 (z=0.163), consists of a pair of bright bipolar-jets that extend over 0.5\,Mpc 
in the LOw-Frequency ARray \citep[LOFAR;][]{vanhaarlem13} DR2 \citep{shimwell22} map. It was found with the upgraded 
Giant Metrewave Radio Telescope \citep[uGMRT;][]{gupta17}. 
Similarly to their senior cousins, the bi-polar radio jets exhibit strong wiggles and edge-darkening. 

Nevertheless, some properties of the newly discovered TRG significantly differ from both TRGs that had been identified previously. The two northern jets of 3C\,75 are seen to merge into a single entity, while still being relatively close to the core region.
A similar situation is likely observed in the other TRG, PKS\,2149$-$158; however, even the most detailed map available lacks 
sufficient clarity, probably due to unfavourable projection effects \citep{guidetti08}.
Secondly, in 3C\,75, all four jets undergo a sharp bending towards east/northeast \citep{owen95}, 
which is attributed to dynamic pressure exerted by an intracluster crosswind
blowing at a speed of 1200 $\rm km \ s^{-1}$ \citep{hudson06}, resulting from an ongoing merger of the host cluster 
Abell\,400 with a sub-cluster \citep[e.g.][]{beers92, hudson06}. This extraneous factor complicates 
analytical modelling and numerical simulations of the basic phenomenon of twin bi-polar jets \citep[e.g.][]{molnar17}.
A similar structural complexity is found in the twin bi-polar jets of the second TRG, PKS\,2149$-$158, 
which exhibits a conspicuous C-shaped bending \citep{guidetti08}, also indicating a strong crosswind acting on the jets.

In contrast, the two nearly parallel wiggly jets of TRG\,J104455+354055 bounce sideways against each other but escape disruption 
for the initial few tens of kpc \citep[see Fig. 1 in][]{gopal-krishna22}.
This level of resilience exhibited by the jets suggests that their magnetic fields might play an important role, 
preserving the identity of the individual jets. While a change in the direction of the magnetic field in a wiggling jet 
had been previously observed \citep[e.g.][]{wezgowiec24}, very little is known about the interactions of two magnetised jets.

Furthermore, the impact of the orbital motion of two host galaxies of TRG\,J104455+354055 on the morphology of the jets 
is not contaminated by the effect of ram-pressure of a cross-wind. This makes this object a much cleaner test case 
and a suitable laboratory for studying the nature of such rare systems.

To study the propagation and interactions of the jets of TRG\,J104455+354055, we performed polarimetric observations 
using the VLA in the extended B and C configurations in the C and X frequency bands. 
Additionally, B-configuration S-band observations were performed, enabling a direct link between our high-frequency data and the data presented in \citet{gopal-krishna22}. We also analysed our VLA data for the WAT source, also reported by \citet{gopal-krishna22}. Due to its angular proximity to TRG\,J104455+354055, this source (WAT\,J104442+354017), 
associated with the galaxy 2MASX\,J10444188+3540170, 
was shown to be present within the primary beam of our S-band and C-band observations. 
Throughout the paper, for the overlays, we used the optical image from the Hyper Suprime-Cam Legacy Archive \citep{tanaka21} and we refer to this image as the HSC image.

In Sect.~\ref{obsred}, we present the details of observations and data reduction. 
Section~\ref{results} presents the obtained results and Sect.~\ref{disc} offers a brief discussion of the results. 
We summarise our findings in Sect.~\ref{conclusions}. Throughout the paper, we assume a flat Universe with $H_0$ = 70 $\rm km\ s^{-1}\ Mpc^{-1}$, $\Omega_m$ = 0.3, and $\Omega_\Lambda$= 0.7.

\section{Observations and data reduction}
\label{obsred}

This paper is based on observations performed by us under the VLA project 24A-019 in January, April, and May 2024.
Because of the relatively small angular extent of the target source (around 15\arcsec), it was observed in the C band and X band in more extended B and C configurations. All these setups enable resolutions comparable to or higher than that of the 
uGMRT observations~\citep[see Table~\ref{vlaobs} and][]{gopal-krishna22}. 
Additionally, S-band B-configuration observations were performed within the same project.
The details of the observations are presented in Table~\ref{vlaobs}.

\begin{table*}[ht]
        \caption{\label{vlaobs}VLA observations of TRG\,J104455+354055.}
\centering
\begin{tabular}{cccccr}
\hline\hline
Band & Central Frequency & Bandwidth & Array & Synthesized       & Time             \\
     & [MHz]             & [MHz]     &       &beamwidth [arcsec] & on source [mins] \\
\hline
S    & 2000              & 2024      & B     & 2.1               & 9                \\
C    & 6000              & 4048      & B     & 1.0               & 130              \\
C    & 6000              & 4048      & C     & 3.5               & 28               \\
X    & 10000             & 4048      & B     & 0.6               & 160              \\
X    & 10000             & 4048      & C     & 2.1               & 60               \\
        \hline
\end{tabular}
\end{table*}

The data were reduced with the Common Astronomy Software Applications (CASA) package\footnote{http://casa.nrao.edu} \citep{mcmullin07}.
All data affected by shadowed antennas or the radio frequency interference (RFI) were flagged manually.
In all observations, the radio sources 3C\,286 and J1033+3935 were used for the flux density scale and phase calibration, respectively.
To account for the polarisation leakage of the antennas, typically of the order of 5\% of the signal, 
a single scan of the radio source J1407+2827 (OQ208) was performed. This source is very bright (around 1\,Jy or more) 
and unpolarised at the frequencies considered in our study. Under such conditions, it is possible to reduce the impact of the instrumental polarisation to 0.1\%\footnote{https://science.nrao.edu/facilities/vla/docs/manuals/cal}. The calibration of the polarisation angle was carried out using observations of 3C\,286.

The data reduction resulted in five calibrated datasets for the C band and X band in both the B and C arrays, as well as the S band in the B array, each containing information about three Stokes parameters I, Q, and U.
Because our observations are focussed on obtaining both high resolution and sensitivity to diffuse structures and the polarised emission, the data from both the B and C arrays were combined in the imaging process for the C and X bands.
To deconvolve the synthesised beam of the interferometer from the dirty map, we used CASA task {\sc tclean} to run the Clark CLEAN algorithm \citep{clark80}.

The use of two different arrays required various observing times to match the sensitivity of the final maps. However, a direct use of both datasets
(B and C array) leads to an increased weighting towards the smaller spatial scales (i.e. more data in the B array observations). We accounted for that by using
{\sc tclean} with {\sc robust=1}, which shifts the weights towards the larger spatial scales. This allowed us to optimally utilise the high resolution of the B-array observations with sensitivity to large-scale emission afforded by the C-array data, which was especially important for Stokes Q and U. The resolutions and sensitivities of the final maps are presented in Table~\ref{finalmaps}. 

The polarisation vectors in all maps presented in this paper are the observed E-vectors rotated by 90\degr. With the assumption of an optically thin emitting region (justified 
in the case of jets and lobes of radio galaxies) the orientation
of the observed polarisation vectors corresponds fully to the orientation of the magnetic fields where the emission is known to originate from \citep[e.g.][]{cawthorne13, wardle18}. 
Furthermore, our observations at frequencies 3\,GHz and higher are scarcely affected by the Faraday rotation effects. For these reasons, we can assume that the observed orientation of the polarisation vectors aptly describes the orientation of the magnetic fields within the source projected on the sky plane.

\begin{table}[ht]
        \caption{\label{finalmaps}Resolutions and sensitivities of the final maps.}
\centering
\begin{tabular}{ccc}
\hline\hline
Band & FWHM       & TP/PI noise level          \\
     & [arcsec]   & [$\mu$Jy/beam]             \\
\hline
S    & 2.0        & 16.7/15.7                  \\
C    & 1.6        & 2.7/2.5                    \\
X    & 1.1        & 2.5/2.1                    \\
\hline
\end{tabular}
        \tablefoot{
                TP -- total power; PI -- polarised intensity.\\ 
        }
\end{table}

\section{Results}
\label{results}

In the analysis of the radio maps of TRG\,J104455+354055 presented in this paper, we follow the convention used in \citet{gopal-krishna22} and 
refer to the host galaxies as A (left) and B (right). Consequently, each of the four jets is named after the host galaxy and the direction of propagation. For example, the southern jet of the right host galaxy is called the BS-jet. All flux densities, both of total (TP) and polarised (PI) radio intensity are presented in Table~\ref{fluxes}.

For a more detailed analysis of the complexity of TRG\,J104455+354055, we applied Sobel filtering to the total intensity radio maps. This filtering is available within the task {\sc niner} of the AIPS package\footnote{http://aips.nrao.edu}. The Sobel operator computes the square root of the sum of the squares of the directional derivatives in two orthogonal directions, which means it is sensitive to edges in the original image. This property is especially useful in following the changes in direction of the radio jets (narrow lines) or finding the limits of bright areas.

\begin{table*}[ht]
        \caption{\label{fluxes}Integrated radio flux densities of TRG\,J104455+354055 and WAT\,J104442+354017.}
\centering
\begin{tabular}{lrrrrrr}
\hline\hline
Source          & \multicolumn{2}{c}{S band}    & \multicolumn{2}{c}{C band}    & \multicolumn{2}{c}{X band}      \\
                & TP [mJy]      & PI [mJy]      & TP [mJy]      & PI [mJy]      & TP [mJy]        & PI [mJy]      \\
\hline
A-core          & 2.2$\pm$0.1   & 57$\pm$17\tablefootmark{a}    & 2.0$\pm$0.1   & 46$\pm$4\tablefootmark{a}       & 2.1$\pm$0.1   & 42$\pm$4\tablefootmark{a}     \\
B-core          & 1.6$\pm$0.1 & 5$\pm$18\tablefootmark{a}       & 1.5$\pm$0.1 & 10$\pm$3\tablefootmark{a}     & 1.5$\pm$0.1 & 11$\pm$3\tablefootmark{a}     \\
AN-jet          & 17.2$\pm$0.5& 2.3$\pm$0.1     & 11.9$\pm$0.4  & 1.7$\pm$0.1   & 8.0$\pm$0.2     & 1.3$\pm$0.1   \\
BN-jet          & 11.2$\pm$0.4& 0.9$\pm$0.1     & 8.1$\pm$0.2   & 0.9$\pm$0.1   & 4.6$\pm$0.1         & 0.5$\pm$0.1 \\
AS-jet          & 12.8$\pm$0.4& 1.8$\pm$0.2     & 9.3$\pm$0.3   & 1.8$\pm$0.1   & 5.3$\pm$0.2     & 1.5$\pm$0.1   \\
BS-jet          & 7.6$\pm$0.3 & 0.7$\pm$0.1     & 5.5$\pm$0.2   & 0.7$\pm$0.1         & 3.3$\pm$0.1   & 0.6$\pm$0.1 \\
A total         & 32.8$\pm$1.0& 4.2$\pm$0.2     & 23.7$\pm$0.7  & 3.6$\pm$0.2   & 15.9$\pm$0.5    & 2.9$\pm$0.2   \\
B total         & 20.4$\pm$0.6& 1.4$\pm$0.1     & 14.8$\pm$0.5  & 1.6$\pm$0.1   & 9.5$\pm$0.3     & 1.2$\pm$0.1   \\
TRG             & 54.5$\pm$1.7& 5.9$\pm$0.3     & 39.0$\pm$1.2  & 5.2$\pm$0.3   & 26.0$\pm$0.8    & 4.2$\pm$0.2   \\
WAT-core        & 0.7$\pm$0.1 & 11$\pm$21\tablefootmark{a}      & 0.5$\pm$0.1   & 0.5$\pm$3.4\tablefootmark{a}& --                & --            \\
WAT E-jet       & 4.8$\pm$0.2 & 1.5$\pm$0.2     & 3.9$\pm$0.1   & 0.9$\pm$0.1         & --            & --            \\      
WAT W-jet       & 4.5$\pm$0.2 & 0.9$\pm$0.2     & 3.2$\pm$0.1   & 0.7$\pm$0.1   & --              & --            \\
WAT             & 10.7$\pm$0.4& 2.6$\pm$0.2     & 8.0$\pm$0.2   & 1.7$\pm$0.1   & --              & --            \\
\hline
\end{tabular}
        \tablefoot{
                TP -- total power; PI -- polarised intensity.\\
                \tablefoottext{a}{flux in $\mu$Jy.}
        }
\end{table*}

\subsection{S band}
\label{sbandres}

\begin{figure*}[ht]
   \centering
        \resizebox{0.37\hsize}{!}{\includegraphics[clip]{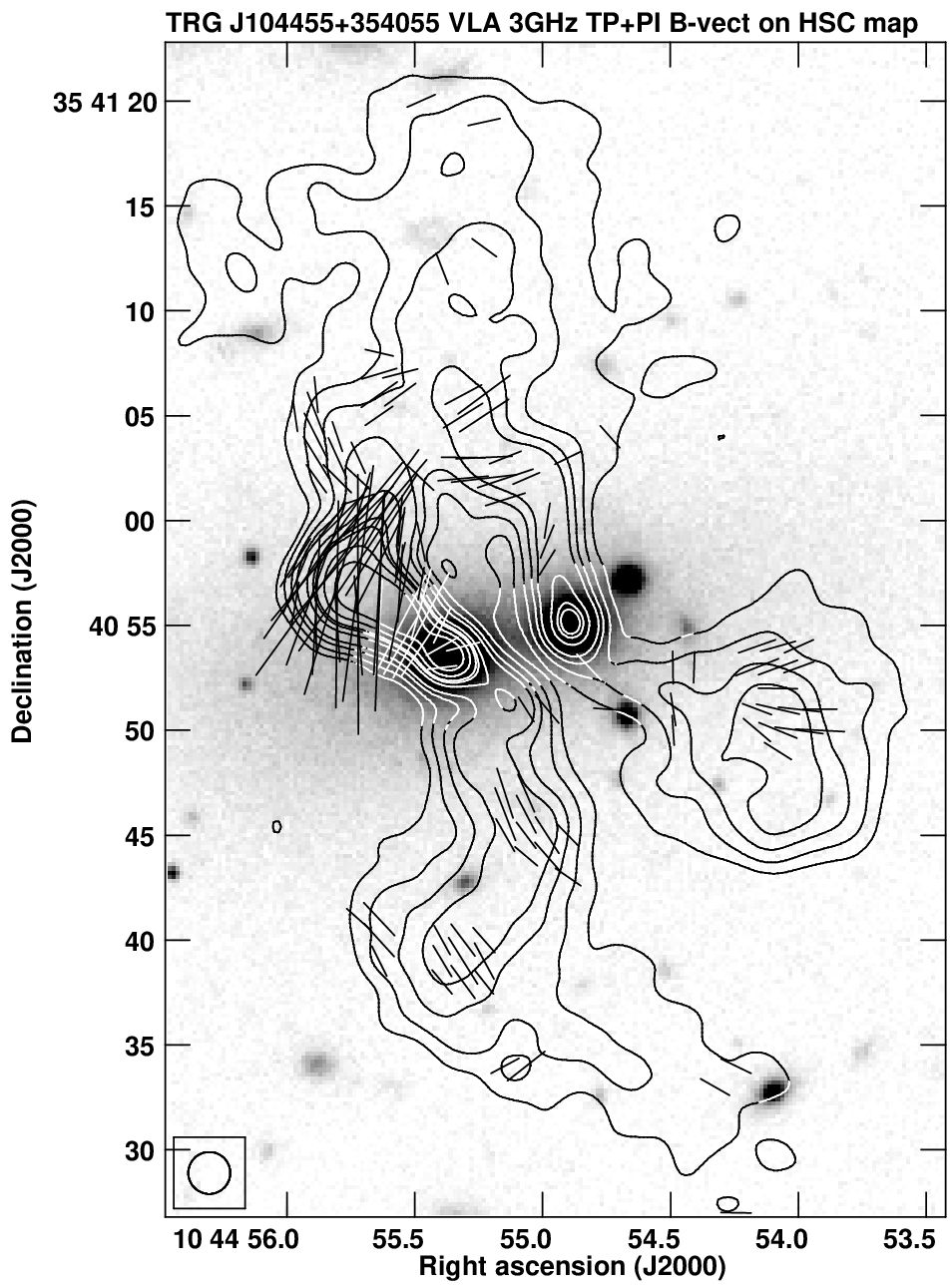}}
        \resizebox{0.61\hsize}{!}{\includegraphics[clip]{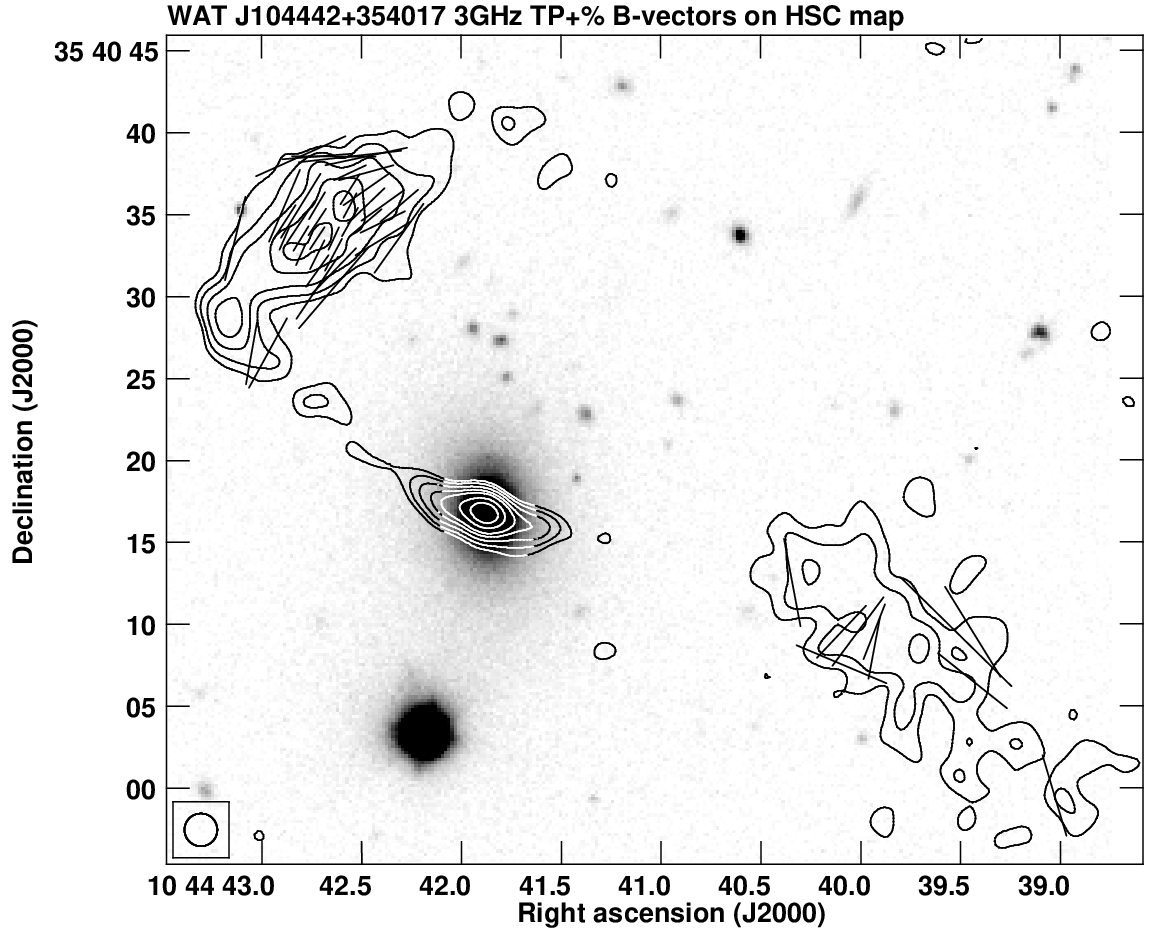}}
        \caption{3\,GHz VLA image of TRG\,J104455+354055 (left) and WAT\,J104442+354017 (right) overlaid on the HSC image. The contour levels
        are 3, 8, 16, 25, 48, 64, 72, 96, 128, 256, 512 (left) and 3, 5, 8, 12, 16, 24, and 32 (right) $\times$ the r.m.s. of 17\,$\mu$Jy/beam. The vectors mark 
        the orientation of the magnetic field in the sky plane. The beam size of 2\arcsec\ is shown in the bottom left corner of each map and the vector lengths equating beam size correspond to the polarised intensity of 67\,$\mu$Jy (left) or the polarisation degree of 25\% (right).}
         \label{sband}
   \end{figure*}

The large-scale radio map of TRG\,J104455+354055 at 3\,GHz (left panel of Fig.~\ref{sband}) 
corresponds well to that observed at 1.4\,GHz by \citet{gopal-krishna22}. The resolution of  
2\arcsec\ allowed us to trace all four jets from the host galaxies up to areas where they decelerate and diffuse into plumes. 
We note here that the largest angular scale at the used B-configuration is 58\arcsec\ in full synthesis 
observations\footnote{https://science.nrao.edu/facilities/vla/docs/\\manuals/oss/performance/resolution}.
It is therefore likely that the moderate time coverage of our observations (see Table~\ref{vlaobs}) 
has led to a significant diminution on this spatial scale in our map and the fainter large-scale emission was not detected. Also, the polarised emission seems to be detected only for the brightest parts of the source. Still, the first sharp bending of the AN-jet is accompanied by strongly polarised emission (with the degree of polarisation reaching 30\%) and the orientation of the B-field vectors hints at a possible compression occurring in this region.

Interestingly, we observe significant differences in the polarised emission from the cores of the two host galaxies (Table~\ref{fluxes}). 
While there is some polarised emission visible in the A-core, the B-core is practically unpolarised. This, however, might be caused by the beam depolarisation of weak signals at lower frequencies, especially given the radio features related to source B are weaker than those related to source A.

The radio map of the neighbouring WAT\,J104442+354017 is also similar to that obtained by \citet{gopal-krishna22} at 1.4\,GHz. Nevertheless, the smaller primary beam 
of the VLA at 3\,GHz means that the western lobe of the radio source falls close to the edge of the primary beam; hence, it was observed with substantially reduced sensitivity, making it difficult to follow the morphology of this lobe. The eastern lobe, however, was imaged with good sensitivity and it is roughly perpendicular to the main axis of the source, which suggests an interaction with a high-velocity flow in the intracluster medium (ICM). The clear alignment of the magnetic field vectors with the bent lobe also seems to confirm this. The core shows very little polarised emission and given the uncertainties, it can be assumed to be unpolarised (Table~\ref{fluxes}).

\subsection{C band}
\label{cbandres}

\begin{figure*}[ht]
   \centering
        \resizebox{0.47\hsize}{!}{\includegraphics[clip]{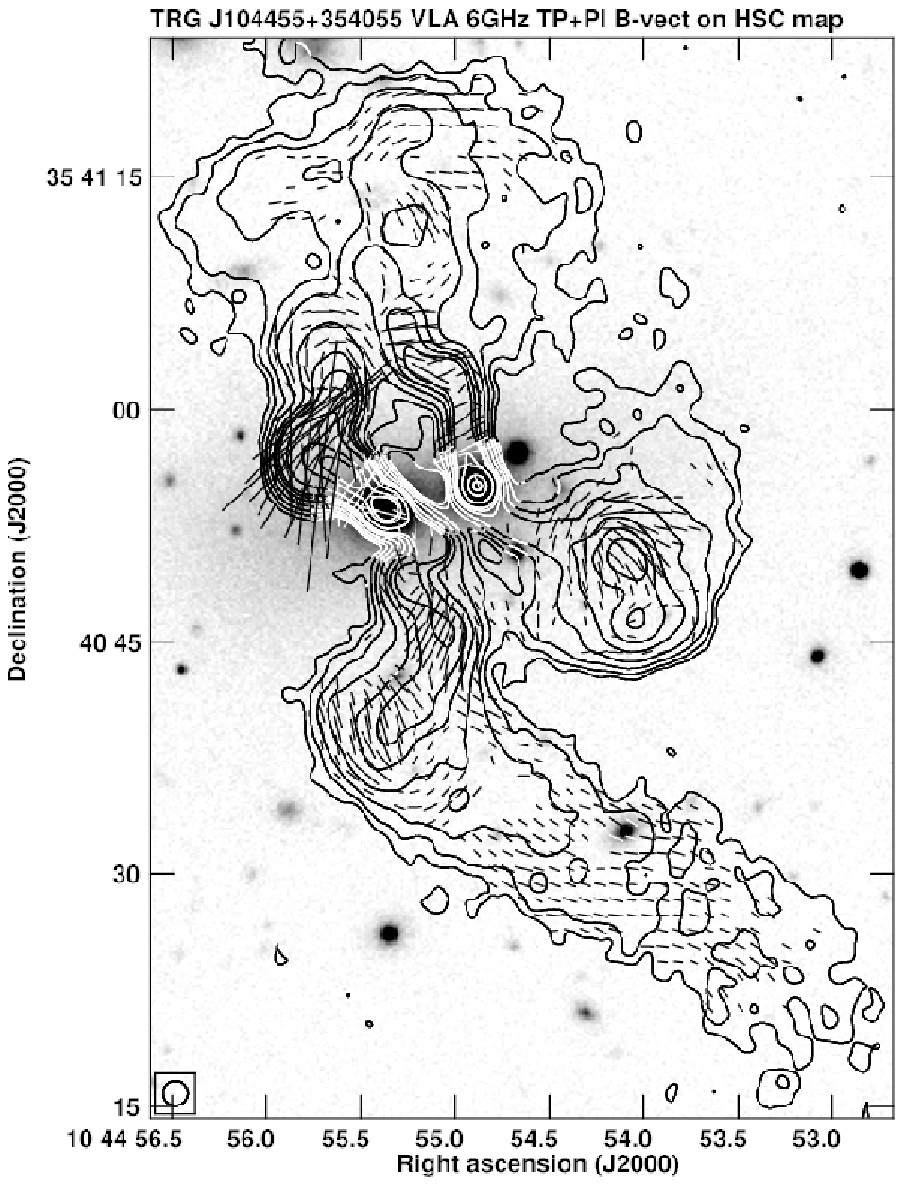}}
        \resizebox{0.47\hsize}{!}{\includegraphics[clip]{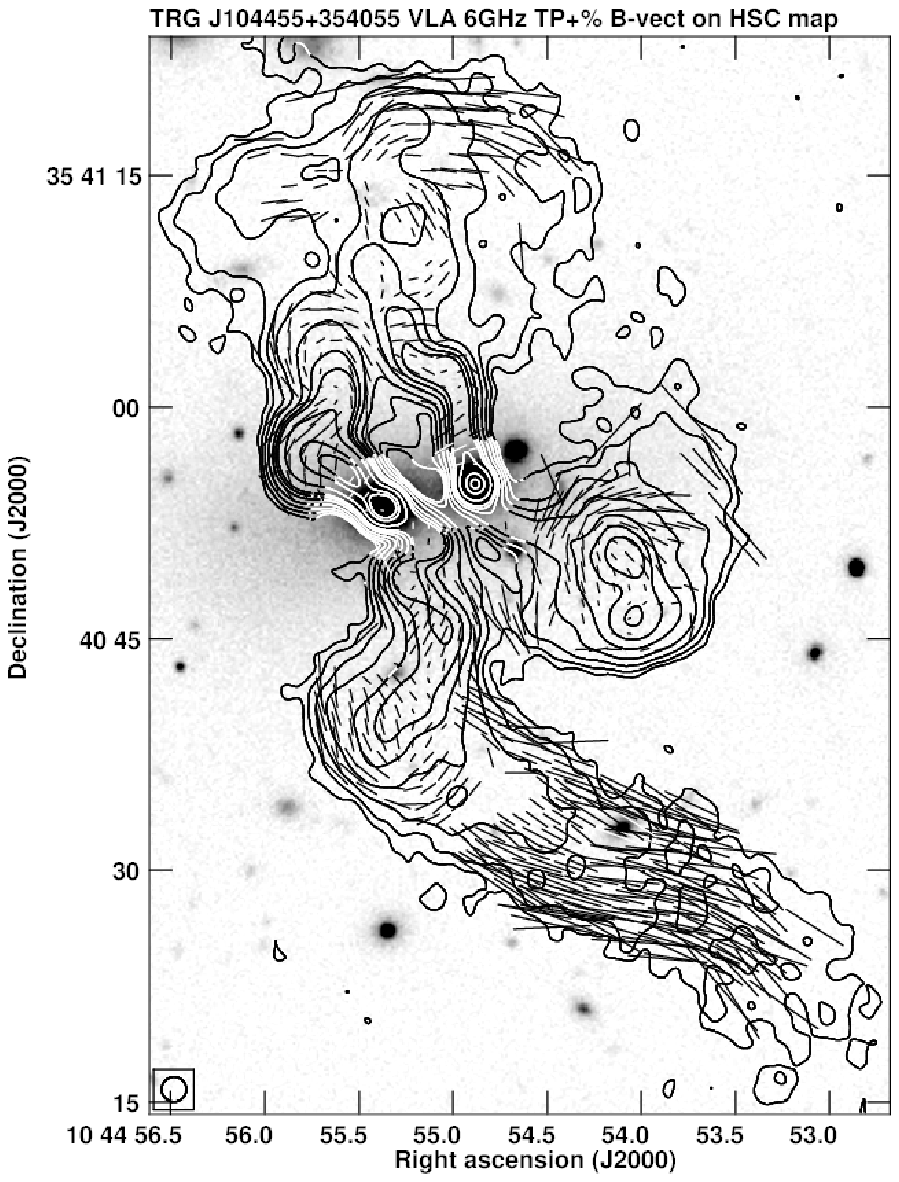}}
        \caption{6\,GHz VLA image of TRG\,J104455+354055 overlaid on the HSC image. The contour levels 
        are 3, 6, 12, 24, 48, 72, 96, 192, 384, and 512\,$\times$\,the r.m.s. of 2.7\,$\mu$Jy/beam. The vectors mark 
        the orientation of the magnetic field in the sky plane. The beam size of 1\farcs6\ is shown in the bottom left corner of each map and 
        the vector lengths equating beam size correspond to the polarised intensity of 27\,$\mu$Jy (left) or the polarisation degree of 27\% (right).}
         \label{cband}
   \end{figure*}

The combination of data from both the B and C configurations of the VLA has yielded a high-resolution map sensitive to large-scale diffuse emission. This is especially visible in the AS-jet, where an impressive and strongly polarised tail-like structure is clearly detected (Fig.~\ref{cband}). 
The degree of polarisation of this structure locally exceeds 50\%.
In contrast, the BS-jet is much more confined and resembles a bubble-shaped lobe with a warm spot. However, the steep intensity gradient observed near the apparent southern edge of this lobe might not necessarily rule out a further radio extension, considering the low polarisation observed in this region. In fact, there is a hint that this lobe even exhibits a plume-like extension, although it most likely extends along the line of sight; hence, it is not clearly distinguishable from the lobe.

\begin{figure*}[ht]
   \centering
        \resizebox{0.48\hsize}{!}{\includegraphics[clip]{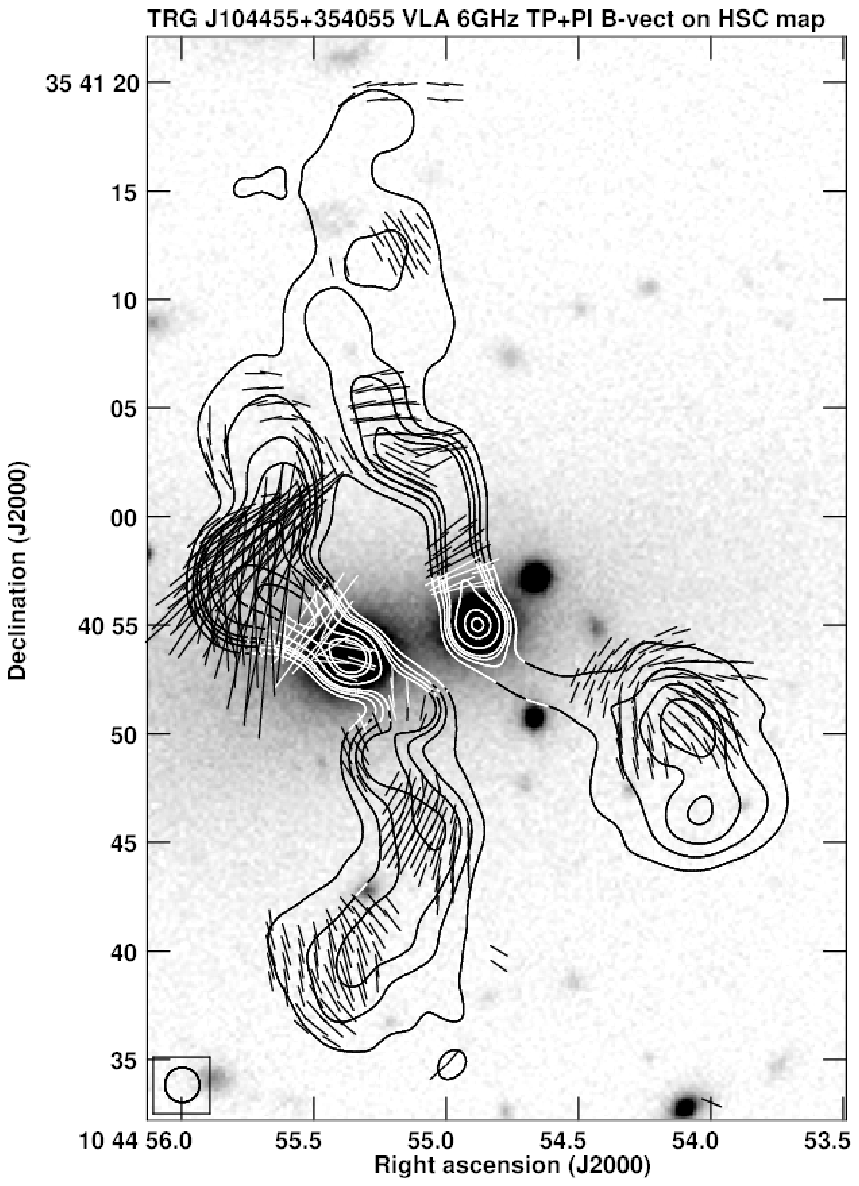}}
        \resizebox{0.46\hsize}{!}{\includegraphics[clip]{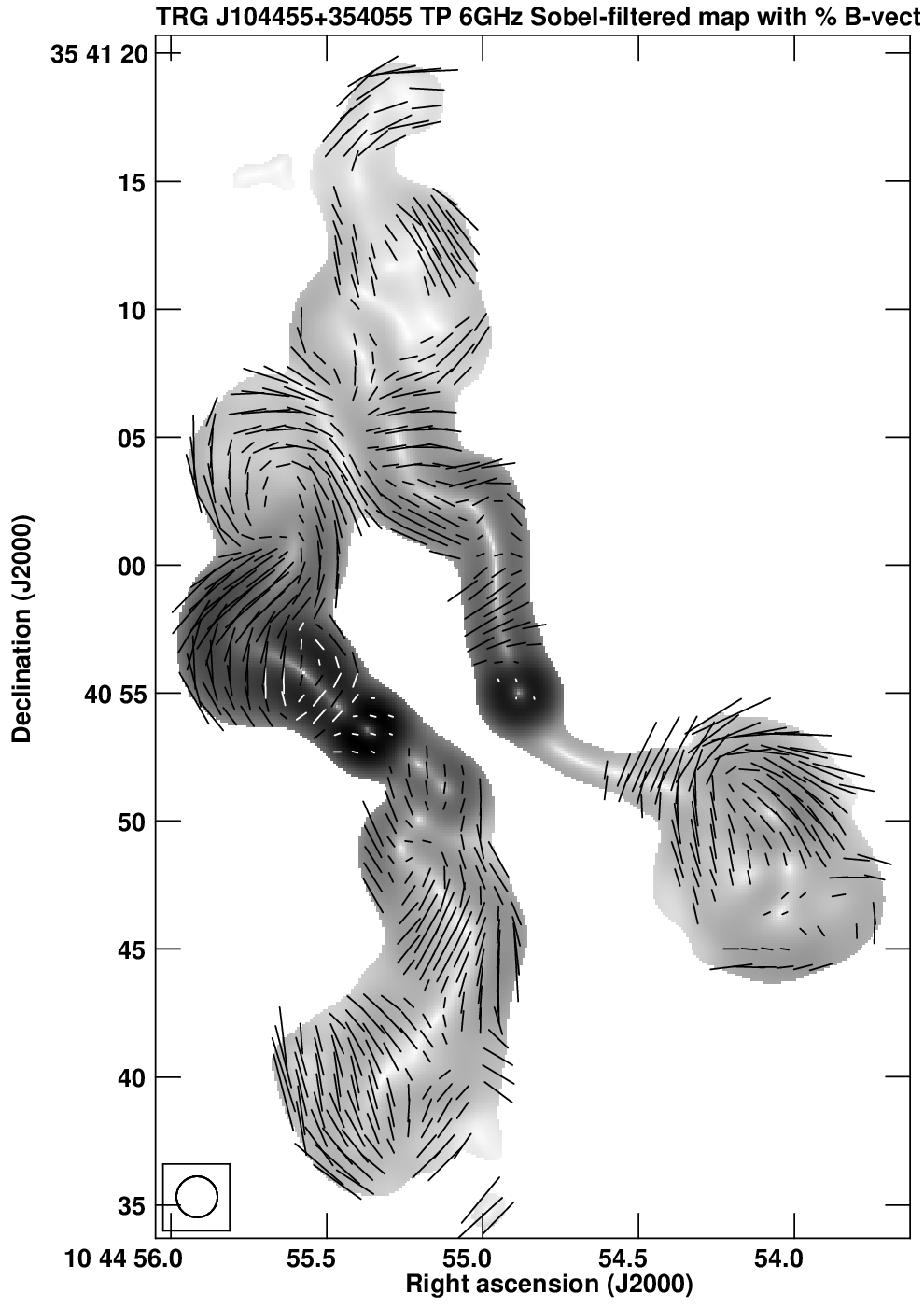}}
        \caption{6\,GHz VLA image of TRG\,J104455+354055 overlaid on the HSC image (left) and the same map with the Sobel filter applied (right). 
        The contours in the left image are the same as in Fig.~\ref{cband} but start from the 24$\sigma$ level. The map on the right is also truncated at this level. The vectors mark the orientation of the magnetic field in the sky plane.     
        The beam size of 1\farcs6\ is shown in the bottom left corner of each map and the vector lengths equating beam size correspond 
        to the polarised intensity of 33\,$\mu$Jy (left) or the polarisation degree of 20\% (right).}
         \label{cbandsobel}
   \end{figure*}

A close-up view of the brightest radio emission allows us to precisely follow all four jets, especially in the Sobel-filtered map (right panel of Fig~\ref{cbandsobel}). 
We observe that while the northern jets of both galaxies bend towards each other, they do not merge or cause any other type of disruption. 
The retention of their identity is probably ensured by the magnetic field oriented along 
the edges of both jets, which prevents the two plasma streams from coalescing. We note that in this case, the orientation of the magnetic field direction within the jets 
changes at the stronger bends from perpendicular to parallel to jet axis. Another example of such a pattern is the giant radio galaxy 
4C\,70.19 \citep{wezgowiec24}. Similarly to the S-band map, the strongest polarised emission 
is observed at the first bend of the AN-jet (with the degree of polarisation again reaching 30\%). The BS-jet shows a very interesting 
orientation of the magnetic field (Fig.~\ref{cbandsobel}). As it is almost unpolarised in its initial narrow part, after the deceleration, a significant polarised emission appears with the magnetic field vectors encompassing the bubble-like structure. As in the S-band map, we measured higher flux densities of the radio features of source A (Table~\ref{fluxes}). Likewise, the polarised emission is much weaker in source B, which suggests that it is an intrinsic property of this source and not a result of low sensitivity to weak emission in the S-band map. We note here that the higher sensitivity of the C-band map enabled the detection of the polarised emission from the core of source B.

\begin{figure*}[ht]
   \centering
        \resizebox{0.48\hsize}{!}{\includegraphics[clip]{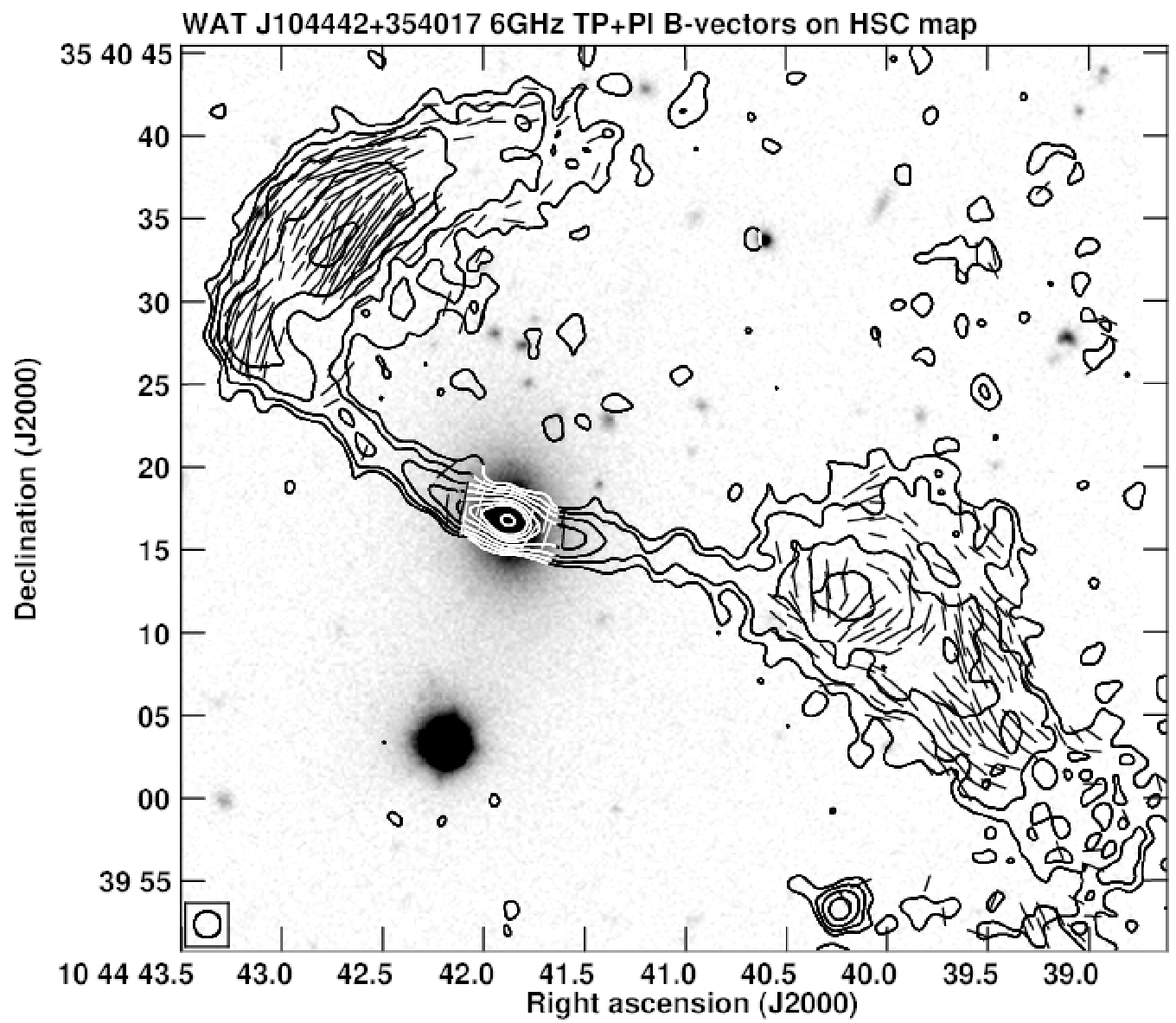}}
        \resizebox{0.48\hsize}{!}{\includegraphics[clip]{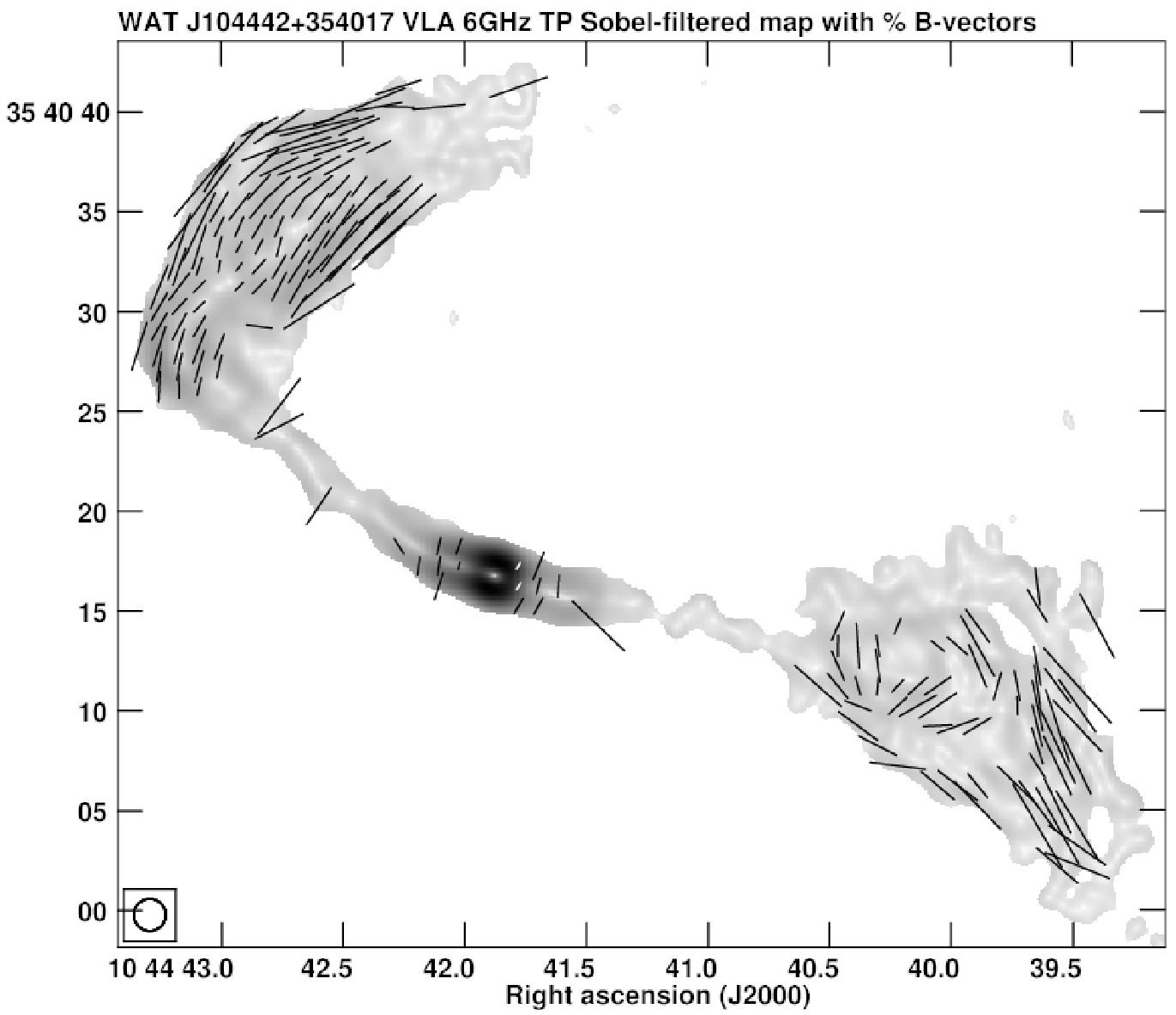}}
        \caption{6\,GHz VLA image of WAT\,J104442+354017 overlaid on the HSC image (left) and the same map with the Sobel filter applied (right). 
        The contours levels in the left image are 3, 6, 12, 24, 48, 72, 96, 192, 384, and 512\,$\times$\,the r.m.s. of 2.7\,$\mu$Jy/beam. The right map is truncated at the 6$\sigma$ level. The vectors mark the orientation of the magnetic field in the sky plane. The beam size of 1\farcs6\ is shown in the bottom left corner of each map and the vector lengths equating beam size correspond to the polarised intensity of 13\,$\mu$Jy (left) or the polarisation degree of 27\% (right).}
         \label{watc}
   \end{figure*}

The higher sensitivity of the C-band map is particularly revealing in the case of WAT\,J104442+354017 (Fig.~\ref{watc}). Both jets can be traced from the core region to the extended plumes. While the core region remains unpolarised, the polarised signal is detected just outside of it, likely where the radio jets emerge from the denser halo of the host galaxy. The orientation of the magnetic field is perpendicular to the jet axis but flips where the eastern jet bends, following the plume extension. The morphology of the western plume, as well as the orientation of its magnetic field, is poorly determined. This is likely the result of the degraded sensitivity close to the edge of the primary beam. 

\subsection{X band}
\label{xbandres}

\begin{figure*}[ht]
   \centering
        \resizebox{0.48\hsize}{!}{\includegraphics[clip]{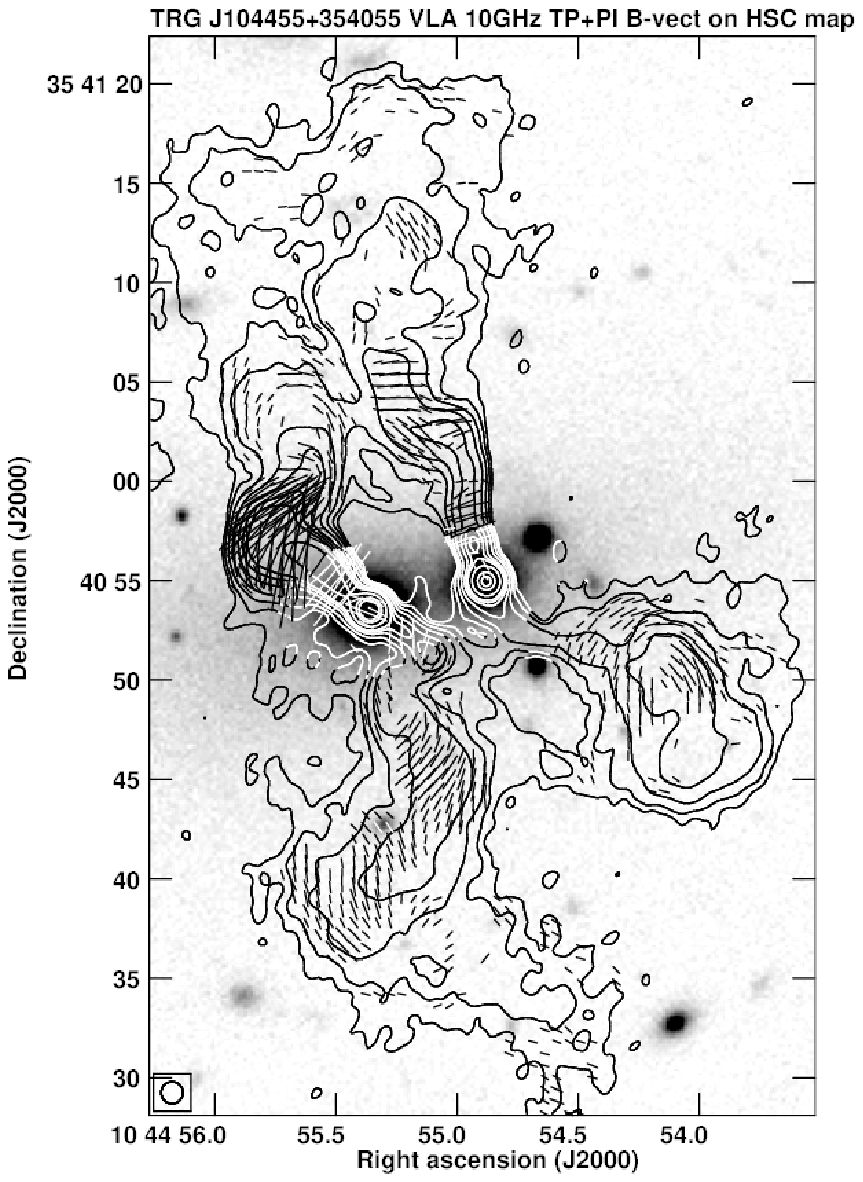}}
        \resizebox{0.51\hsize}{!}{\includegraphics[clip]{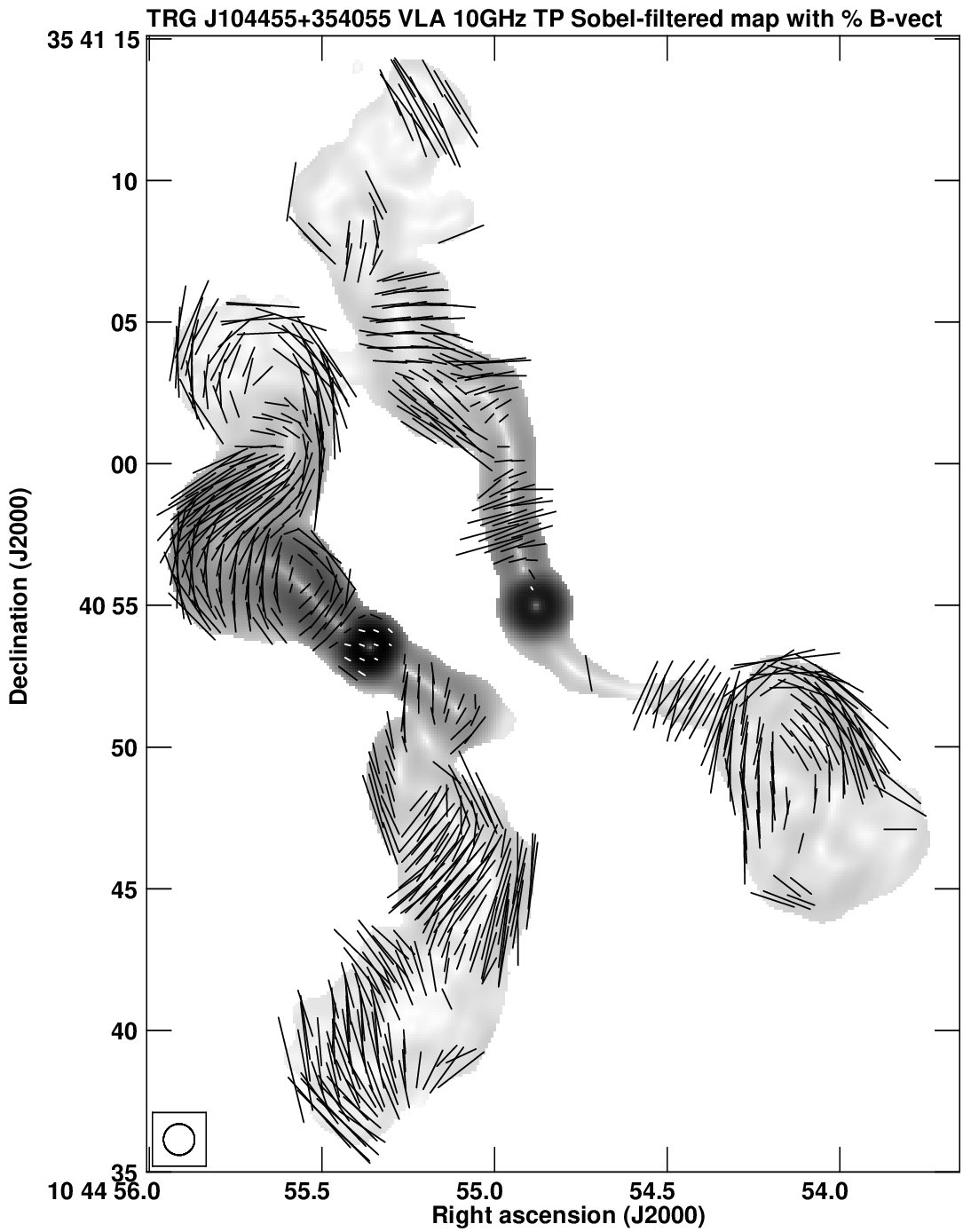}}
        \caption{10\,GHz VLA image of TRG\,J104455+354055 overlaid on the HSC image (left) and the same map with the Sobel filter applied (right). The contour levels
        in the left image are 3, 6, 12, 24, 48, 72, 96, 192, 384, and 512\,$\times$\,the r.m.s. of 2.5\,$\mu$Jy/beam. The map on the right is truncated at the 12$\sigma$ level. The vectors mark the orientation of the magnetic field in the sky plane. 
        The beam size of 1\farcs1\ is shown in the bottom left corner of each map and the vector lengths equating beam size correspond
        to the polarised intensity of 18\,$\mu$Jy (left) or the polarisation degree of 14\% (right).}
         \label{xband}
   \end{figure*}

The overall extent of TRG\,J104455+354055 in the X-band map is roughly the same as that visible in the map at the much lower frequency of the S band (left panel of Fig.~\ref{xband}). This is likely a combined outcome of the limited sensitivity of B-configuration S-band observations (as already discussed in Sect.~\ref{sbandres}) 
and the high sensitivity of the X-band observations, similar to that seen for the C band (see Table~\ref{finalmaps}). At the same time, the absence of the tail-like structure in the AS-jet, as seen in C-band maps (see Sect.~\ref{cbandres}), can be explained with the expected steep spectral index 
of the radio emission from the lobes. In this sense, the emission visible at the 3 to 5$\sigma$ level in the C-band map falls below the detection limit at the X band. This aspect is further discussed in Sect.~\ref{spixes}.

The remaining radio features in the X-band map of TRG\,J104455+354055 correspond well with the C-band map, but the higher resolution of the former clearly demonstrates that the orientation of the magnetic field in the A-core is parallel to the jet axis (Fig.~\ref{xband}) and changes rapidly to the perpendicular 
orientation right outside the core region (and likely the dense halo of the host galaxy). A similar correspondence with the C-band map is observed for source B with the unpolarised BS-jet in its initial narrow part. It seems to extend beyond the gaseous halo of the host galaxy and, thus, it could suggest that this jet is oriented outwards with regard to the sky plane; namely, it would extend beyond the host galaxy. The orientation of the jets of TRG\,J104455+354055 is discussed in Sect.~\ref{disc}. 

\section{Discussion}
\label{disc}

\subsection{Spectral index}
\label{spixes}

\begin{table*}[ht]
        \caption{\label{lffluxes}Low-frequency total radio flux densities of TRG\,J104455+354055 and WAT\,J104442+354017.}
\centering
\begin{tabular}{lrrr}
\hline\hline
Source                  & 144\,MHz\tablefootmark{a}     & 325\,MHz\tablefootmark{b}         & 1.4\,GHz\tablefootmark{c}     \\
                        & [mJy]                         & [mJy]                         & [mJy]                   \\
\hline
TRG\,J104455+354055     & 590$\pm$31                    & 333$\pm$67                     & 108$\pm$22                    \\
WAT\,J104442+354017     & 128$\pm$12                    & 85$\pm$17                      & 23$\pm$2                      \\
\hline
\end{tabular}
        \tablefoot{
                \tablefoottext{a}{LOFAR Two-metre Sky Survey \citet{shimwell22}}
                \tablefoottext{b}{Westerbork Northern Sky Survey \citet{rengelink97}}
                \tablefoottext{c}{NRAO VLA Sky Survey \citet{condon98}}
                }
\end{table*}

\begin{figure*}[ht]
   \centering
        \resizebox{0.49\hsize}{!}{\includegraphics[clip,angle=-90]{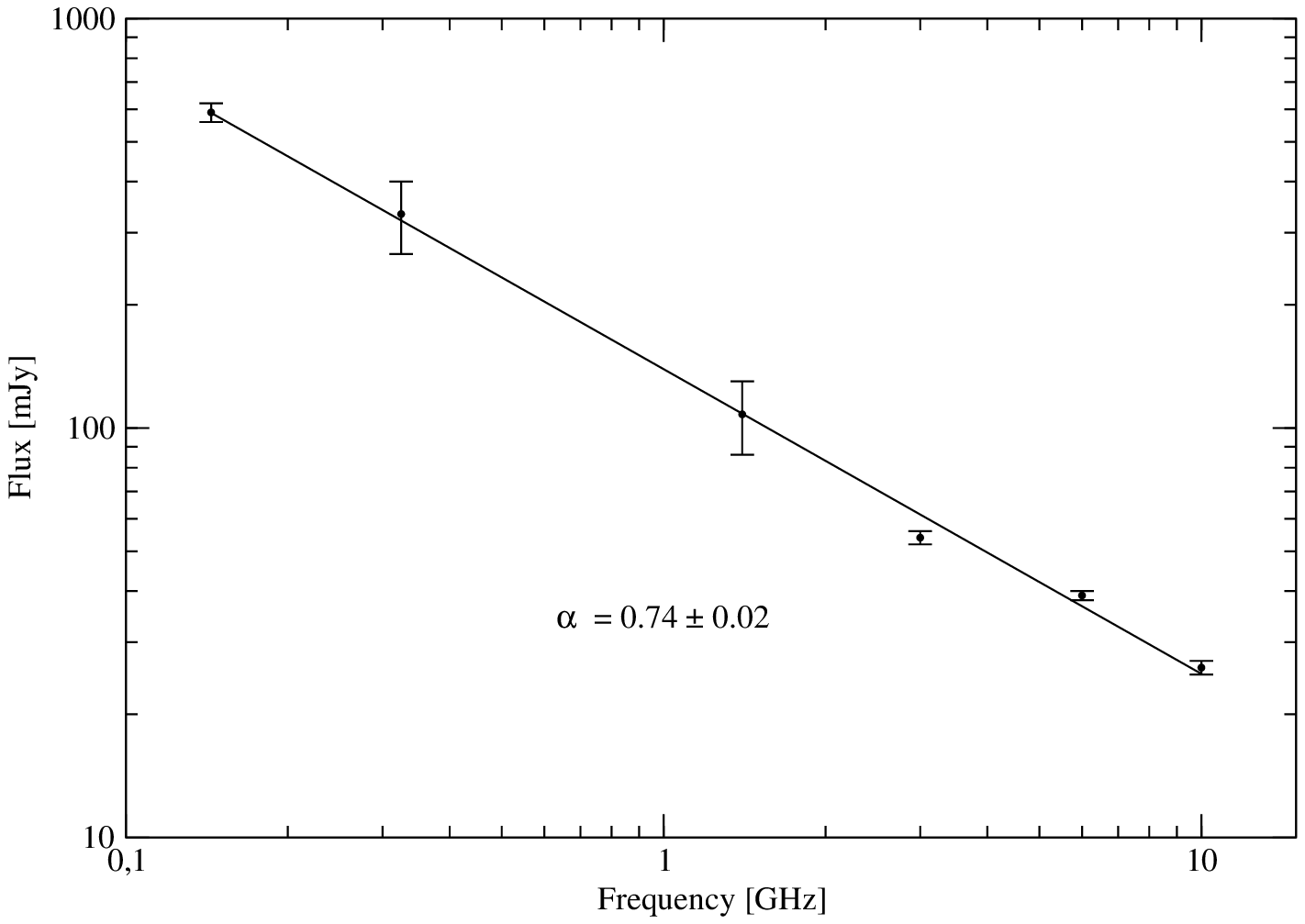}}
        \resizebox{0.49\hsize}{!}{\includegraphics[clip,angle=-90]{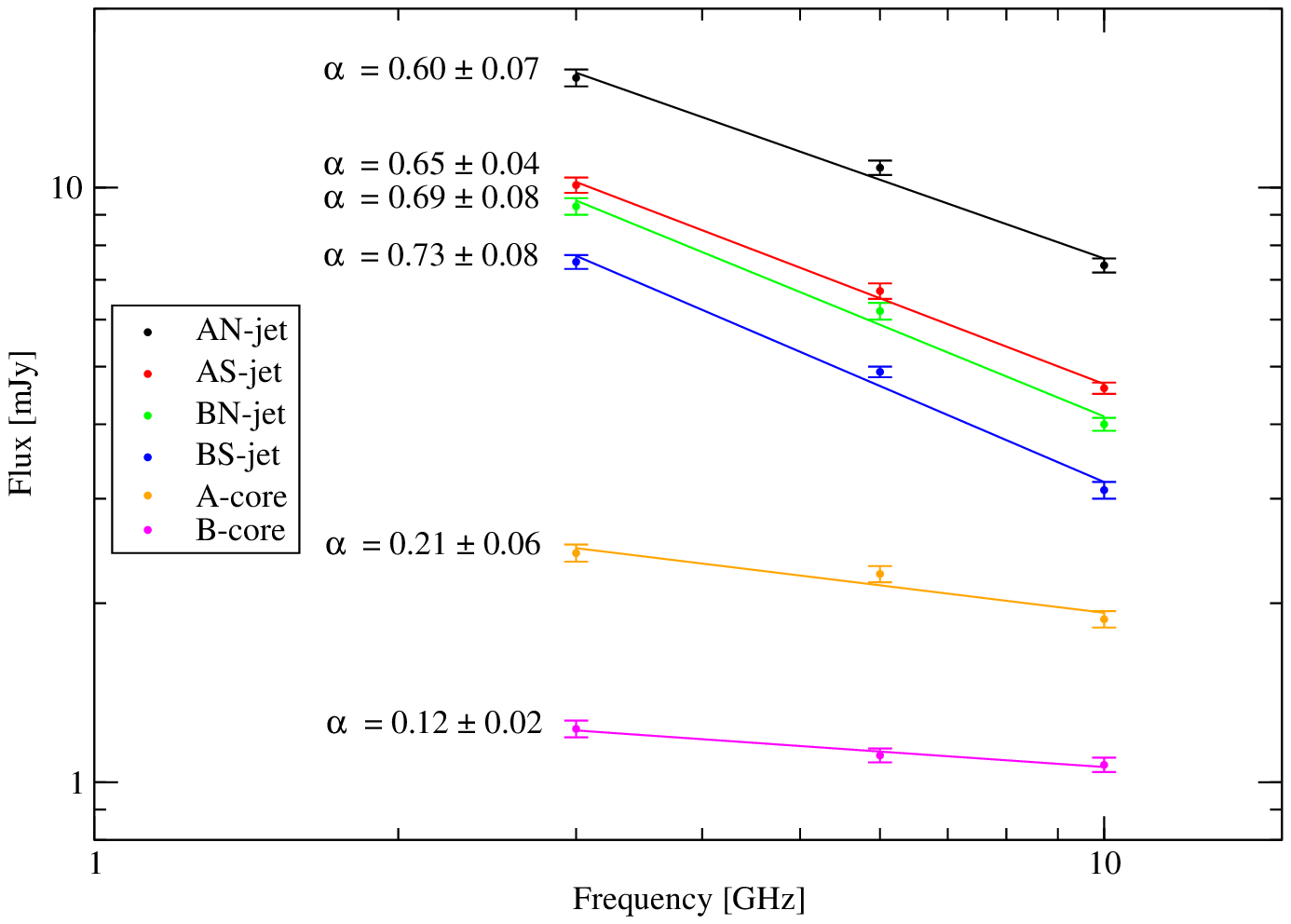}}
        \caption{Global (left) and separate components (right) spectra of TRG\,J104455+354055 with the corresponding values of the spectral 
        index obtained from the linear fit to the spectra. The details are provided in Sect.~\ref{spixes}. 
        }
         \label{spixplot}
\end{figure*}

The total flux densities of TRG\,J104455+354055 derived from our observations, together with the low-frequency flux 
densities used by \citet{gopal-krishna22} 
allowed us to construct a global spectral index of this source over a wide range of frequencies 
(see Table~\ref{lffluxes} and the left panel of Fig.~\ref{spixplot}).
The obtained spectral index of 0.74$\pm$0.02\footnote{$S\propto\nu^{-\alpha}$} 
perfectly matches the result reported by \citet{gopal-krishna22}. We note here that the flux 
at 3\,GHz found below the fit is likely due to the use of the B-configuration data (as mentioned in Sect.~\ref{sbandres}). The missing $\sim$10\,\% (as expected from the fit) should be recoverable using more compact configurations of the VLA. Therefore, it is most likely the case that the spectral index of TRG\,J104455+354055 remains constant between 144\,MHz and 10\,GHz. Nevertheless, 
since these integrated flux densities of the TRG represent the combined emission from two separate radio galaxies, they cannot be used directly to infer their physical properties. Because our high-frequency data allowed us to carry out the mapping with an improved spatial resolution (2\arcsec\ or better), we performed 
a similar fit of the spectral index for the individual components of TRG\,J104455+354055. 
As mentioned before, this includes all four jets and both radio cores. All these spectral fits are presented in a single plot (right panel of Fig.~\ref{spixplot}). Because the plumes from the northern jets merge at larger distance from the host galaxies and the tail-like plume of the AS-jet is visible only in the C-band map, we measured the flux densities of all jets only in the areas where they are robustly discernible. The fitted spectral indices are given in Fig.~\ref{spixplot}.

\begin{figure}[ht]
   \centering
        \resizebox{\hsize}{!}{\includegraphics[clip,angle=-90]{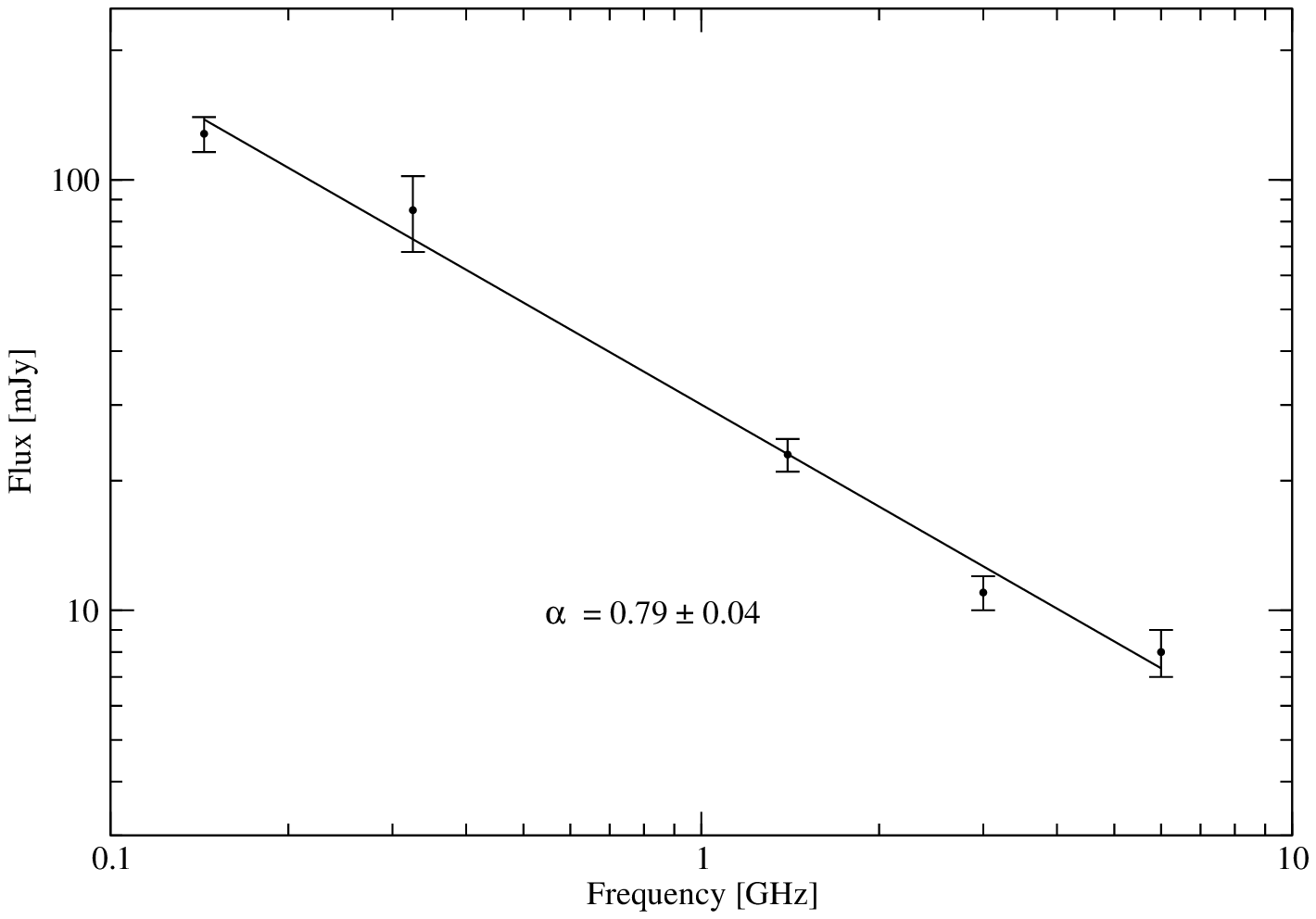}}
        \caption{Global spectrum of WAT\,J104442+354017 with the corresponding value of the spectral
        index obtained from the linear fit to the spectra. The details are provided in Sect.~\ref{spixes}.
        }
         \label{watspix}
\end{figure}

For WAT\,J104442+354017, it was possible to construct only the global spectral index, done using the 
same procedure as for TRG\,J104455+354055. We measured the flux density at 325\,MHz, which was not provided by \citet{gopal-krishna22}, directly in the map from the Westerbork Northern Sky Survey \citep[WENSS;][]{rengelink97}. 
The resulting global spectral index of 0.79$\pm$0.04 (Fig.~\ref{watspix}) almost equals that measured for TRG\,J104455+354055. The slightly higher flux density at 325\,MHz likely results from a large beam of the observations (54\arcsec$\times$92\arcsec) that 
includes a nearby background radio source visible to the north-west from WAT\,J104442+354017 
in the map from the Faint Images of the Radio Sky at Twenty-cm survey \citep[FIRST;][]{becker95}.
On the other hand, the flux density measured in the S-band map is expected to be lower (as mentioned above) due to the use of the more extended B-configuration of the VLA.

\subsection{Magnetic fields}
\label{magfields}

\begin{figure}[ht]
   \centering
        \resizebox{\hsize}{!}{\includegraphics[clip]{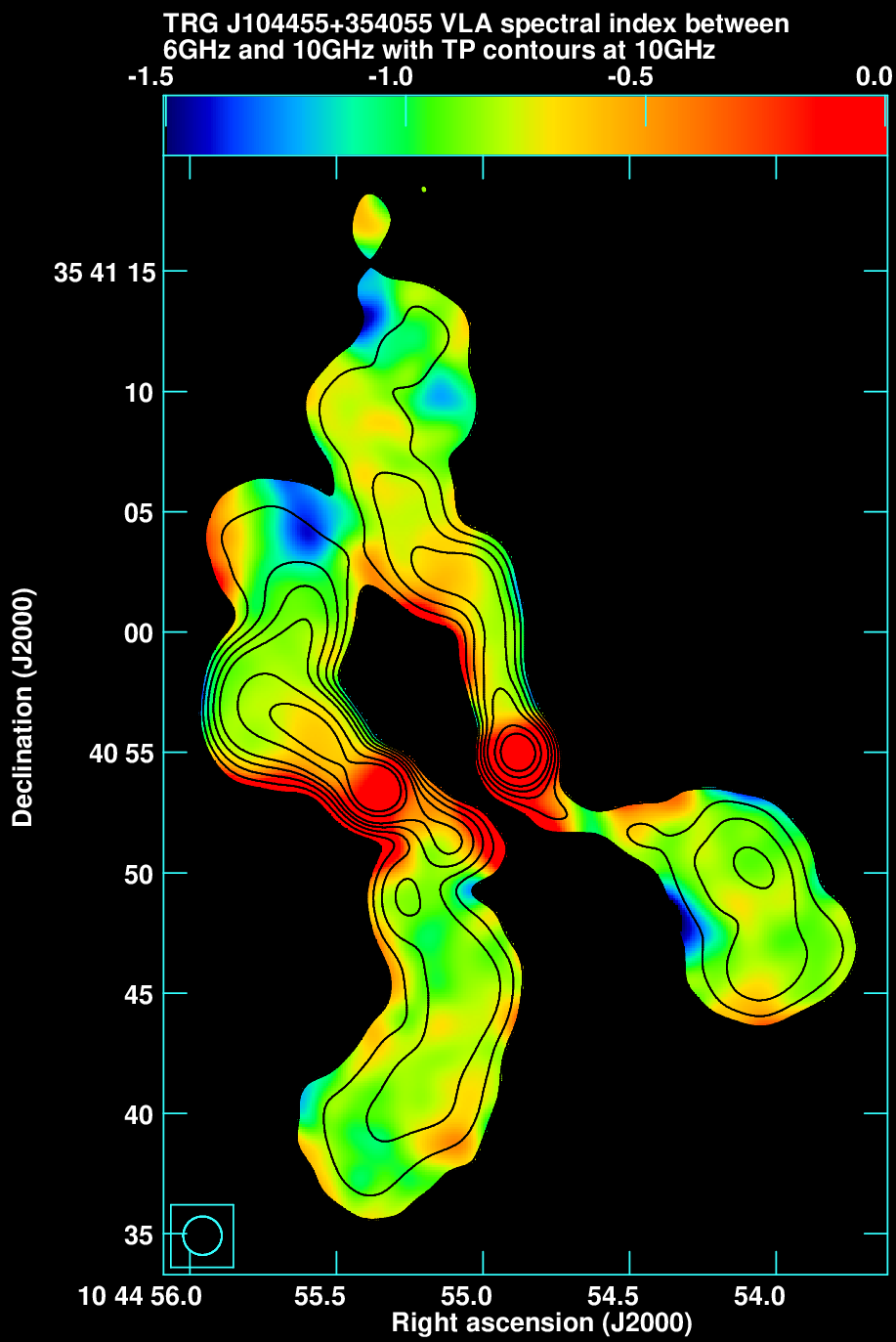}}
        \caption{Spectral index map of TRG\,J104455+354055 between 6 and 10\,GHz overlaid with the 10\,GHz contours.
        Both image and contours start from the 13$\sigma$ level of the 10\,GHz map.}
         \label{spix}
   \end{figure}

\begin{figure}[ht]
   \centering
        \resizebox{\hsize}{!}{\includegraphics[clip]{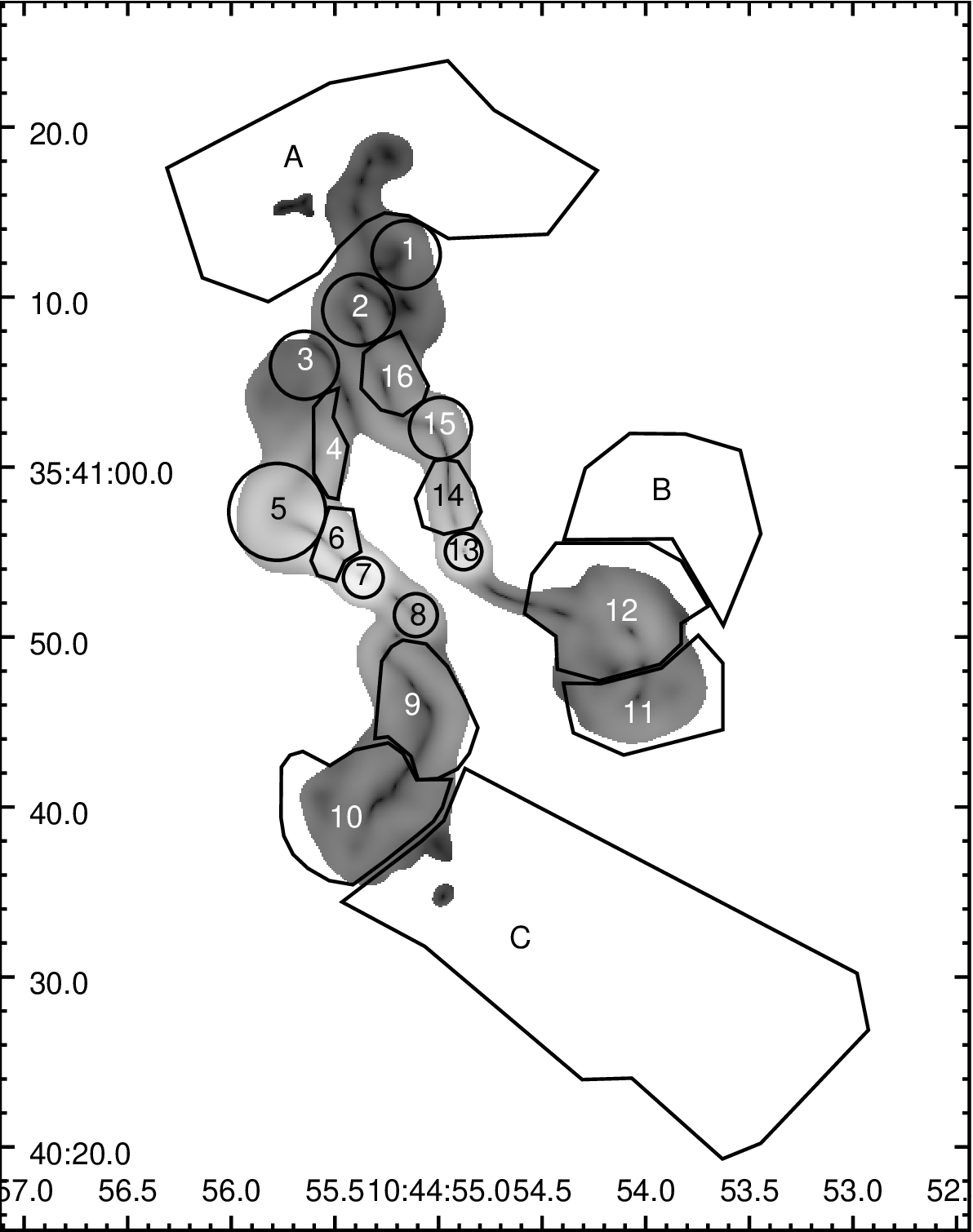}}
        \caption{Regions of TRG\,J104455+354055 used to calculate the magnetic field strengths overlaid on the C-band map with the Sobel filter applied        from the right panel of Fig.~\ref{cbandsobel}. See Sect.~\ref{magfields} for details.}
         \label{bregsT}
   \end{figure}

The polarised emission detected in our sensitive observations of jets and plumes of TRG\,J104455+354055 suggests that the magnetic fields may play an important role in the propagation and interactions of the jets with each other and with the surrounding environment. We used the flux densities derived from the C-band maps (of both total and polarised emission) to calculate the energies and strengths of the magnetic fields in TRG\,J104455+354055. To perform a more detailed analysis, we constructed 
a map of spectral index between C and X bands (Fig.~\ref{spix}) that helped us to select regions of the jets 
with roughly uniform spectral indices. The map was truncated at the 13$\sigma$ level of the X-band map to assure 
a high signal-to-noise ratio and consequently shows only the bright parts of the jets of TRG\,J104455+354055. 
The regions used for our analysis are presented in Fig.~\ref{bregsT}. We also included the three regions of the plumes
at the ends of all the jets (regions A, B, and C), which are visible in the C-band map (Fig.~\ref{cband}). We also adopted a spectral index of 1 for each region in the calculations described below. 

\begin{figure}[ht]
   \centering
        \resizebox{\hsize}{!}{\includegraphics[clip]{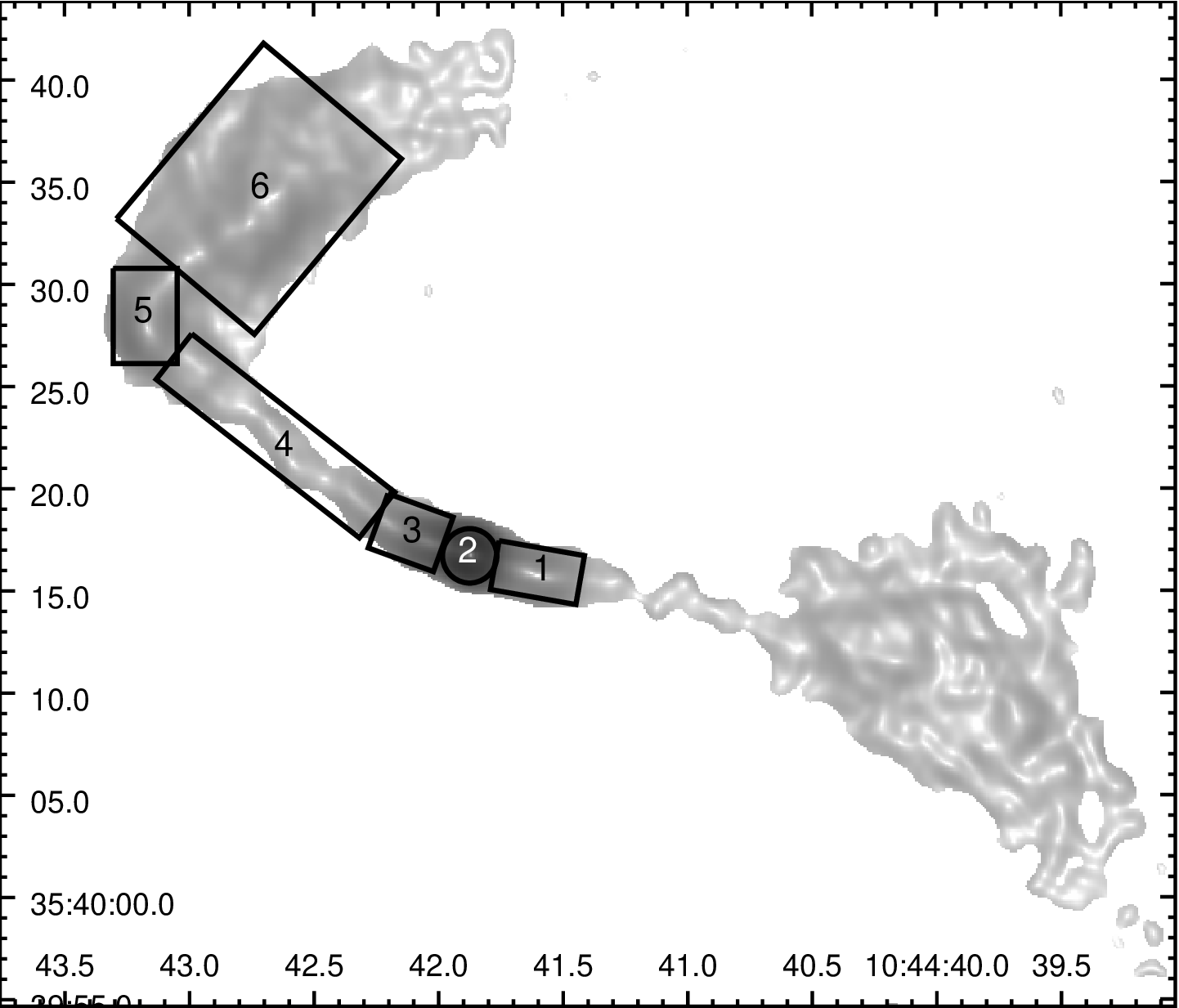}}
        \caption{Regions of WAT\,J104442+354017 used to calculate the magnetic field strengths overlaid on the C-band map with the Sobel filter applied from the right panel of Fig.~\ref{watc}. See Sect.~\ref{magfields} for details.}
         \label{bregsW}
   \end{figure}

A similar analysis was performed for WAT\,J104442+354017, also using the 
C-band maps to derive flux densities and the spectral index map between S-band and C-band for a crude estimate of the spectral index
for each of the selected region (Fig.~\ref{bregsW}).

For the estimations of energies and strengths of the magnetic field, of both the total and ordered component, 
we used the web application {\sc bmag}\footnote{https://bmag.oa.uj.edu.pl} based on the calculations 
performed by \citet{zychowicz25}. The details are presented in Appendix A. 
The derived values of the strengths of the magnetic field, both total and ordered, 
as well as energy densities of the total magnetic field in all regions shown in Figs.~\ref{bregsT} and~\ref{bregsW} 
are presented in Table~\ref{magparams}. We applied the Sobel filter to the C-band maps in Figs.~\ref{bregsT} and~\ref{bregsW} to highlight the directions of propagation of the jets.

As expected, the highest strengths of the total magnetic field are observed for the radio cores A and B (regions 7 and 13, respectively). Interestingly, similarly high values were obtained for the AN-jet (regions 3 through 6) with the magnetic field slightly stronger in its innermost part (region 6) than in the A-core. We note here, however, that the strength of the magnetic field in both cores are likely 
the lower limits because we might expect the presence of protons in the halos of host galaxies. 
In such cases, the strengths would increase to around 15\,$\mu$G 
for a proton-to-electron ratio of 1 and to around 25\,$\mu$G for a ratio of 10.

A relatively strong magnetic field in the western side of the AN-jet (regions 3 and 4) might be explained by the likely amplification
of the magnetic field, as suggested by the magnetic field vectors in the right panel of Fig.~\ref{cbandsobel} and 
relatively high strengths of the ordered magnetic field (Table~\ref{magparams}).
Such envelope of the magnetic field around the AN-jet could be also the reason for its better collimation when compared to the BN-jet. 
We discuss this further in Sect.~\ref{jetprop}. 
Although weaker strengths of the magnetic field were derived for both southern jets compared to their northern counterparts, 
similar values of around 4\,$\mu$G are found in areas where the jets decelerate and expand into plumes (regions A, B, and C).

Due to the off-axis position of WAT\,J104442+354017
(with respect to the centre of the primary beam), a reliable analysis was possible only for its eastern lobe and the core region. 
As expected, the highest strength of the magnetic field was derived for the core. Of course, this value should be considered 
to be a lower limit, as argued previously in the case of the host galaxies of TRG\,J104455+354055. 
In contrast to those galaxies, however, the host galaxy of 
the WAT\,J104442+354017 ejects radio jets that are nearly perfectly symmetric. Their bases (regions 1 and 3) are characterised with the same strengths of both total and ordered magnetic fields.
In the straight part of the eastern jet, the strengths of the total and 
the ordered magnetic field are 4.9$\pm$0.4\,$\mu$G and 2.0$^{+0.2}_{-0.1}$\,$\mu$G, respectively. In the region where the jet turns by around 
90\degr\,, the strength of the total magnetic field increases only slightly (within the uncertainties); however, that of the ordered component 
increases by almost a factor of 2. This is likely due to the shearing of the plasma flow and the associated 
magnetic field which changes its orientation from perpendicular 
to parallel to the jet in the region of the bend, as visible in the right panel of Fig.~\ref{watc}. After the bend, the jet decelerates rapidly into a plume, whereby the strength of the total magnetic field is 4.1$\pm$0.3\,$\mu$G. This corresponds very well to similar regions 
of TRG\,J104455+354055.

\begin{table}[ht]
        \caption{\label{magparams} Properties of the magnetic fields in TRG\,J104455+354055 and WAT\,J104442+354017.}
    \centering
\begin{tabular}{ccccc}
\hline\hline
             &          &       B$_{tot}$       &$\epsilon_B$           & B$_{ord}$               \\
Reg.         &$\alpha$  &       [$\mu$G]        &[10$^{-12}$            &[$\mu$G]                 \\
             &          &                       &erg\,cm$^{-3}$]        &                         \\
\hline
\vspace{5pt}
T1           & 0.88     &  7.2$\pm$0.6          & 2.1$^{+0.4}_{-0.3}$   &  4.0$\pm$0.3            \\
\vspace{5pt}
T2           & 0.80     &  6.9$^{+0.6}_{-0.5}$  & 1.9$\pm$0.3           &  2.1$^{+0.2}_{-0.1}$  \\
\vspace{5pt}
T3           & 1.24     &  9.9$\pm$0.9          & 4.0$^{+1.1}_{-0.8}$   &  4.7$\pm$0.4            \\
\vspace{5pt}
T4           & 0.94     &  10.4$^{+0.9}_{-0.8}$ & 4.3$^{+0.8}_{-0.6}$   &  4.7$^{+0.4}_{-0.3}$    \\
\vspace{5pt}
T5           & 0.87     &  10.3$^{+0.8}_{-0.7}$ & 4.2$^{+0.7}_{-0.6}$   &  5.9$^{+0.5}_{-0.4}$    \\
\vspace{5pt}
T6           & 0.68     & 11.6$^{+0.6}_{-0.5}$  & 5.4$^{+0.6}_{-0.5}$   &  4.4$^{+0.3}_{-0.2}$    \\
\vspace{5pt}
T7           & 0.21     & 11.4$^{+0.3}_{-0.5}$  & 5.2$\pm$0.3           &  3.1$\pm$0.2            \\
\vspace{5pt}
T8           & 0.59     & 8.6$\pm$0.4           & 2.9$\pm$0.3           &  3.1$\pm$0.2            \\
\vspace{5pt}
T9           & 0.82     & 7.0$^{+0.8}_{-0.5}$   & 2.0$^{+0.3}_{-0.2}$   &  3.5$^{+0.3}_{-0.2}$    \\
\vspace{5pt}
T10          & 0.76     &  5.8$^{+0.6}_{-0.3}$  & 1.3$\pm$0.2           &  3.1$\pm$0.2            \\
\vspace{5pt}
T11          & 0.84     &  6.6$^{+0.7}_{-0.5}$  & 1.8$\pm$0.3           &  2.1$^{+0.2}_{-0.1}$    \\
\vspace{5pt}
T12          & 0.98     &  7.1$^{+0.8}_{-0.6}$  & 2.0$^{+0.5}_{-0.3}$   &  3.8$^{+0.4}_{-0.3}$    \\
\vspace{5pt}
T13          & 0.12     &  9.9$^{+0.3}_{-0.4}$  & 3.9$^{+0.2}_{-0.3}$   &  2.0$\pm$0.1    \\
\vspace{5pt}
T14          & 0.77     &  8.3$^{+0.6}_{-0.5}$  & 2.7$^{+0.4}_{-0.3}$   &  3.9$^{+0.3}_{-0.2}$    \\
\vspace{5pt}
T15          & 0.86     &  8.9$^{+0.9}_{-0.6}$  & 3.2$^{+0.6}_{-0.4}$   &  3.4$^{+0.3}_{-0.2}$    \\
\vspace{5pt}
T16          & 0.76     &  7.5$\pm$0.5          & 2.2$\pm$0.3           &  3.8$\pm$0.3            \\
\vspace{5pt}
TA           & 1.00     &  4.9$\pm$0.5          & 1.0$\pm$0.2           &  2.8$^{+0.3}_{-0.2}$    \\
\vspace{5pt}
TB           & 1.00     &  4.3$^{+0.5}_{-0.4}$  & 0.8$^{+0.2}_{-0.1}$   &  2.6$^{+0.3}_{-0.2}$    \\
\vspace{5pt}
TC           & 1.00     &  4.1$^{+0.5}_{-0.4}$  & 0.7$^{+0.2}_{-0.1}$   &  3.2$^{+0.4}_{-0.3}$    \\
\vspace{5pt}
W1           & 0.70     &  5.8$\pm$0.4          & 1.4$\pm$0.2           &  2.7$\pm$0.2            \\
\vspace{5pt}
W2           & 0.20     &  6.9$\pm$0.2          & 1.9$\pm$0.1           &  1.0$\pm$0.1            \\
\vspace{5pt}
W3           & 0.6      &  5.7$\pm$0.3          & 1.3$\pm$0.1           &  2.6$^{+0.2}_{-0.1}$    \\
\vspace{5pt}
W4           & 0.8      &  4.9$\pm$0.4          & 1.0$\pm$0.2           &  2.0$^{+0.2}_{-0.1}$    \\
\vspace{5pt}
W5           & 0.6      &  5.4$\pm$0.3          & 1.2$\pm$0.1           &  3.7$\pm$0.2            \\
\vspace{5pt}
W6           & 0.7      &  4.1$\pm$0.3          & 0.7$\pm$0.1           &  2.9$\pm$0.2            \\
\hline
\end{tabular}
\tablefoot{
        The columns are region number (T for TRG\,J104455+354055 and W for WAT\,J104442+354017),
        spectral index, strength, and energy density of total magnetic field, and ordered magnetic field strength.
}
\end{table}

\subsection{Propagation of the jets}
\label{jetprop}

In the case of isolated radio galaxies, the appearance of their jets and lobes depends mainly 
on the power of the central active galactic nucleus (AGN) and the properties of the surrounding intergalactic medium. When studying 
objects that reside in groups or clusters of galaxies, we can expect that the orbital motion of the host galaxy, 
as well as its interactions with other galaxies influence significantly the morphology of the associated 
radio source \citep[e.g.][]{heesen18,wezgowiec24}. Also the precession of the AGN, induced by the interactions, 
can cause complex morphology of the jets and lobes emanating from the host galaxy \citep[e.g.][]{misra25}.

It is clear that we might expect a combination of these effects in TRG\,J104455+354055 associated 
with two elliptical galaxies forming a binary system. Recent simulations of this source 
by \citet{mondal25} who used the total radio emission maps indeed suggest that 
the wiggles and bends of the jets might be attributed to the orbital motion of the two host galaxies.

Our sensitive high-resolution maps of both total and polarised radio intensity open the possibility 
to construct a 3D picture of the jets of both sources presented in this paper and should provide 
important constraints for future more detailed magnetohydrodynamic (MHD) simulations.

To search for such constraints, we analysed the propagation of the jets using the total radio emission maps 
with the Sobel filter applied. We also used the projected orientation of the magnetic field vectors, 
obtained from the polarised radio emission (right panels of Figs.~\ref{cbandsobel},~\ref{watc}, and~\ref{xband}).

The intensity gradients in Sobel-filtered maps allowed us to trace the axes of the jets, with their 
continuity marking the propagation roughly in the sky plane. Any deviation from this direction is visible 
as darker boundaries of the jets. In that sense, the first turn of the AS-jet (region 8 in Fig.~\ref{bregsT}) 
would be outwards and the second (between regions 8 and 9) again towards the observer. Then, a slight bend in region 9 would occur roughly in the sky plane followed by a sharp turn outwards in region 10. The orientation of the magnetic field vectors support such a scenario, because the magnetic field might shear when the jet is bent significantly \citep[as observed in 4C\,70.19, see][]{wezgowiec24}. 
We note here that the magnetic field vectors remain roughly perpendicular to the axis of the AS-jet until region 10, where they change orientation to parallel. The freely expanding plume of 
the AS-jet (region C) is characterised by a strongly ordered magnetic field.

A similar situation is observed in the eastern jet and plume of WAT\,J104442+354017 where a sharp bend of the jet is accompanied by a change in the orientation of the magnetic field (from perpendicular to parallel with regard to the jet axis) and its significant ordering in the plume formed after the bend. Some hints of this behaviour of the magnetic field in jet bends 
can be found in a study of a sample of WAT galaxies \citep{hardcastle04}. In addition, recent MeerKAT \citep{jonas16} observations of another 
WAT galaxy J1712-2435 \citep{cotton25} offer a distinct case of such field re-ordering, which makes it likely a common property of strongly bent jets that is not necessarily found in WAT galaxies, but in any radio galaxies that show a complex morphology of their jets and lobes or plumes. A good case that could be used to characterise such a source could be the aforementioned 4C\,70.19 \citep{wezgowiec24}.

For the BS-jet of TRG\,J104455+354055, we propose that the bubble-like structure (region 12 in Fig.~\ref{bregsT}) is probably not a terminal hot spot, as the total intensity images might suggest. 
It is more likely a projection of the jet that has sharply turned, 
almost directly away from the observer, while the arcing vectors of the magnetic field are simply shown to be changing their direction from perpendicular to parallel to the jet (right panel of Fig.~\ref{xband}). 
Region B, being the diffuse emission visible above the bright part of the BS-jet in Fig.~\ref{cband}, 
would be then a diffuse plume at the end of this jet, which is in agreement with the strength 
of the magnetic field in this area, similar to other plumes (regions A and C).

In the north of TRG\,J104455+354055, the first bend of both jets seems to occur in the position where they 
emerge out of the denser halo of their host galaxies (left panels of Figs.~\ref{cbandsobel} and~\ref{xband}). 
This behaviour of the jets was also observed by \citet{wezgowiec24} 
in 4C\,70.19 and its host galaxy NGC\,6048. Then, the BN-jet turns slightly, preserving the orientation of the 
magnetic field perpendicular to its axis, while the AN-jet sharply turns changing the orientation of its magnetic 
field due to shear (right panels of Figs.~\ref{cbandsobel} and~\ref{xband}). This change likely prevented 
both jets from merging in the area of their so-called close encounter (between regions 3/4 and 2/16 in Fig.~\ref{bregsT}). 
As made visible in the right panels of Figs.~\ref{cbandsobel} and~\ref{xband}, the magnetic field creates an envelope confining the AN-jet. This, along with the much more elongated shape of the BN-jet, suggests that the large-scale plume visible in the 
north of TRG\,J104455+354055 might be primarily associated with the BN-jet. It would also offer a direct evidence of the importance of the magnetic field in the propagation of the radio jets. These findings, together with the derived strengths and energies of the magnetic field, should enable
more precise simulations of such complex, yet rare sources in the future.

\section{Conclusions}
\label{conclusions}

In this paper, we present sensitive a multi-frequency study of a unique system TRG\,J104455+354055 consisting of two elliptical galaxies that are likely interacting. Their AGNs would therefore be producing extended radio jets that interact with each other and with the surrounding intracluster medium. Another radio galaxy, WAT\,J104442+354017, was found within the primary beams of S-band and C-band observations. With the help of the available polarimetric data, we were able to trace the propagation of the jets of the radio sources, as well as study the properties of their magnetic fields. Our main findings are:

\begin{itemize}

        \item[-] The orientation of the magnetic field is perpendicular to the jet axis only until a sharp bend. In the 
                bends, the magnetic field is undergoing shearing, which changes its orientation to parallel to the jet axis. As a result, there is a slight increase in the strength of the magnetic field as well as in the ordering. This is observed for the jets of both TRG\,J104455+354055 and WAT\,J104442+354017. Along with other similar objects found in the literature, it suggests it would be a more general property of the jets of radio galaxies.
        \item[-] A compact feature seen near the edge of the south-western lobe of TRG\,J104455+354055 does not point to the formation of a hot spot; rather, it is due to a projection effect. We are likely observing the outwardly directed bend of the jet. This scenario is supported by the diffuse radio emission that is visible above and beyond the lobe. In addition, the magnetic fields in this area have similar properties to those observed in other plumes of the radio jets.
        \item[-] The two northern jets meet but do not merge, which is likely caused by a relatively strong and ordered parallel magnetic field along both jets, especially the eastern one.

\end{itemize}

The results of our study suggest that magnetic fields play an important role in the evolution of radio sources. Further analyses of the radio polarised emission from radio galaxies are required to provide much needed additional input for future MHD simulations of jet propagation.

\begin{acknowledgements}
We thank the anonymous referee for the comments, that helped to improve the clarity of this paper. 
GK thanks Indian National Science Academy for a Senior Scientist position during which this project was initiated.
LCH was supported by the National Science Foundation of China (12233001) and the China Manned Space Program (CMS-CSST-2025-A09).
This paper is based in part on data from the Hyper Suprime-Cam Legacy Archive (HSCLA), 
which is operated by the Subaru Telescope. The original data in HSCLA was collected 
at the Subaru Telescope and retrieved from the HSC data archive system, 
which is operated by the Subaru Telescope and Astronomy Data Center at National Astronomical Observatory of Japan. 
The Subaru Telescope is honored and grateful for the opportunity of observing the Universe from Maunakea, 
which has the cultural, historical and natural significance in Hawaii.
This paper makes use of software developed for the Vera C. Rubin Observatory. 
We thank the observatory for making their code available as free software at  http://dm.lsst.org.
\end{acknowledgements}

\bibliographystyle{aa}
\bibliography{myreferences.bib}
 
\begin{appendix}
\section{{\sc bmag} web application}

As mentioned in Sect.~\ref{magfields}, to derive the strengths of the total and ordered magnetic fields, as well as their 
energy densities, we used the {\sc bmag} web application that implements the Bayesian approach to calculating equipartition magnetic fields, 
as presented in \citet{zychowicz25}. The application allows us to set several parameters and priors, as follows:

\begin{itemize}
\item[-] configuration of the magnetic field: we calculated both uniform and random components;
\item[-] source geometry: the emitting volumes associated with the segregated individual regions were approximated with an ellipsoid or a horizontal cylinder for regions with circular or irregular appearance, respectively (see Fig.~\ref{bregsT}). For the latter, both dimensions of the cylinder
were calculated to match the area of the corresponding irregular region. We note here that such volume estimates
are fairly crude, especially taking into account the likely complex 3D morphology of the source(s).
However, changing the volume by a factor of 2 would change the computed magnetic field by just 15\%;
\item[-] cosmological corrections and source proximity: we performed calculations for a distant source (z=0.163);
\item[-] radio emission: we set the central frequency of the C band (6\,GHz) and assume that the entire flux density measured in the relevant areas is non-thermal. 
Their synchrotron spectral indices are estimated using the spectral index map presented in Fig.~\ref{spix}. For the flux densities, both total and polarised, as well as for 
spectral indices we adopt uncertainties of 10\%. We also assumed that the average orientation of the uniform magnetic field relative 
to the sky plane is 45\,degrees. This seems to be a reasonable approximation to the configuration of the magnetic field in radio jets that rapidly change their direction.
The volume filling factor is always set to unity;
\item[-] energy spectra of cosmic rays: we used the flat (default) low-energy spectrum without a cutoff of the high-energy spectrum. The latter was needed for regions with a spectral index of around 0.7 or flatter to obtain good quality fits to the data. 
        The threshold was set to 52.6\,GeV, as suggested by \citet{zychowicz25} for such cases;
\item[-] K0: the ratio of proton and electron density: we assumed purely leptonic plasma and set a constant value of 0;
\item[-] magnetic field Bayesian priors: fixed between 1\,$\mu$G and 15\,$\mu$G;
\item[-] Metropolis-Hastings method of the Markov chain Monte Carlo (MCMC) with all parameters set to default was used, as suggested by \citet{zychowicz25}.
\end{itemize}

With the above settings, it was possible to obtain good quality fits to all the spatial regions that were analysed and the resulting parameters of the magnetic field are 
presented in Table~\ref{magparams}.
\end{appendix}
\end{document}